\documentclass[11pt, a4paper, times]{article}

\usepackage[top=1in, bottom=1.25in, left=1.25in, right=1.25in]{geometry}
\usepackage{moreverb}

\usepackage{hyperref}
\usepackage{graphicx}
\usepackage{multirow}
\usepackage{pifont}
\usepackage{listings}
\usepackage{alltt}
\usepackage{float}
\usepackage{placeins}
\usepackage{dblfloatfix}

\usepackage[usenames,dvipsnames,svgnames,table]{xcolor}

\usepackage{czt}
\usepackage{circus}
\usepackage{hijac}

\usepackage{float}

\usepackage{soul}

\makeatletter
\def\doi#1{\gdef\@doi{#1}}\def\@doi{}
\newfont{\mycrnotice}{ptmr8t at 7pt}
\newfont{\myconfname}{ptmri8t at 7pt}
\makeatother

\lstnewenvironment{Java}
{\lstset{numbers=left,basicstyle=\fontsize{8}{10}\ttfamily,xleftmargin=0pt,
        language=Java}}{}
        
\newenvironment{syntax}{\begin{displaymath}\begin{array}{lcl}}{\end{array}\end{displaymath}}

\begin{document}

%\runningheads{Luckcuck, Wellings and Cavalcanti} {Safety-Critical Java: Level 2 in Practice}

\title{Safety-Critical Java: Level 2 in Practice}

\author{Matt Luckcuck, Andy Wellings and Ana Cavalcanti}

\date {16th September 2016}
%\address{Department of Computer Science, University of York, York, YO10 5GH, UK}

%\corraddr{Andy Wellings, Department of Computer Science,  University
%of York, York, YO10 5DD, UK, E-mail: andy.wellings@york.ac.uk}

%\maketitle 
\thispagestyle{empty}
\maketitle
%%%%%%%%%%%%%%%%
%   ABSTRACT   %
%%%%%%%%%%%%%%%%

\begin{abstract}
\noindent
Safety Critical Java (SCJ) is a profile of the Real-Time Specification for Java that brings to the safety-critical industry the possibility
of using Java. SCJ defines three compliance levels: Level 0, Level 1 and Level 2.
The SCJ specification is clear on what constitutes a Level 2 application in terms of its use of the defined API, but not the occasions on which it should be used.
This paper broadly classifies the features that are only available at Level 2 into three groups:~nested mission sequencers, managed threads, and global scheduling across multiple processors.
We explore the first two groups to elicit programming requirements that they support.
We identify several areas where the SCJ specification needs modifications to support these requirements fully;
these include:~support for terminating managed threads,
the ability to set a deadline on the transition between missions, and augmentation of the mission sequencer concept to support
composibility of timing constraints. We also propose simplifications to the termination protocol of missions and their mission sequencers. To illustrate the benefit of our changes, we present excerpts from a formal model of SCJ Level~2 written in \Circus{}, a state-rich process algebra for refinement.

\end{abstract}

\section{Introduction}

An international effort has produced a specification for a high-integrity real-time version of Java:~Safety-Critical Java~(SCJ)~\cite{SCJDraft}. SCJ is based on a subset of Java augmented by the Real-Time Specification for Java~(RTSJ)~\cite{Wellings2004}, which supplements Java's garbage-collected heap memory model with support for memory regions~\cite{tt97} called memory areas.

\begin{figure}[t]\centering
  \includegraphics[scale=0.45]{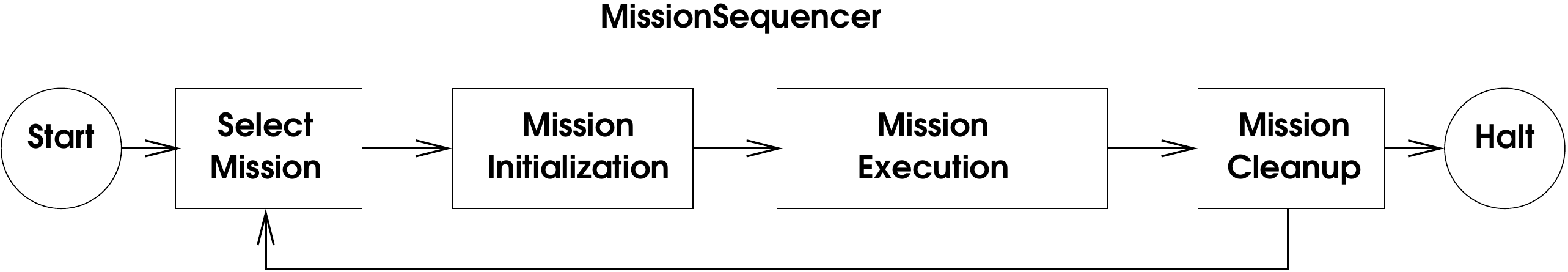}
  \caption{Safety Critical Mission Phases (taken from~\cite{SCJDraft}) \label{phases}}
\end{figure} 

The SCJ programming model is based on the notion of a mission. Each mission consists of a set of periodic (PEH), aperiodic (APEH), and one-shot (OSEH) event handlers, and no-heap managed real-time threads. The execution of a mission progresses through an initialisation, execution, and cleanup phase~(see Figure \ref{phases}). A mission's handlers and threads are created and registered during its initialisation phase. A mission continues to execute until one of its handlers, threads, or a peer mission requests termination, causing control to flow into a mission cleanup phase. An application-defined mission sequencer determines the sequence of missions to be executed.

SCJ restricts the RTSJ memory model to prohibit use of the heap, and defines a policy for the use of the RTSJ's immortal and scoped memory areas. Each managed thread and event handler of the programming model described above has an associated memory area, that holds temporary objects created during execution of that component. When a managed thread terminates, and each time an event handler finishes its handling of and event, the memory area is exited and all of the memory allocated for the component's temporary objects is reclaimed. An immortal memory area holds objects throughout the lifetime of the program:~they are never deallocated. The scoped memory area of a mission is cleared out at the end of each mission. Each release of a handler has an associated per-release scoped memory area, cleared out at the end of the release. In the case of a thread, the execution of its associated \texttt{run()} method is viewed as a single release, and consequently, it is associated with its own local scoped memory area. Additionally, during a release, a stack of temporary private scoped memory areas can be used.

The SCJ language specification defines three compliance levels (Levels 0, 1 and 2), which reflect three supported programming and execution models.
The compliance levels reflect increased levels of complexity in terms of the available programming features and, therefore, of the supported programs, with direct impact on the effort required for certification. It is accepted that the effort required to certify a program that exploits the generality of Level~2 capabilities may be significantly greater than that required to certify programs that use only the more limited capabilities of the lower compliance levels.

The differences between the three compliance levels are summarised in Table~\ref{ComplianceLevels}. The schedulable objects available at each level include those listed for that level and those listed for the previous levels. While mission sequencers are schedulable objects and are used at all compliance levels, they can only be registered to a mission at Level~2, as we explain below. The Suspension column refers to the availability of features like \verb"Object.wait()", \verb"Object.notify()", and \verb"Services.delay()".

\begin{table}[h]
\centering
  \begin{tabular}{ c || l | l| l | l }
    \hline
    		& \parbox[t]{3cm}{Execution\\ Model }					   & \parbox[t]{3cm}{Schedulable\\ Objects }			& Suspension & \parbox[t]{3cm}{Platform} \\ \hline \hline 
Level~0 & Cyclic Executive 					   & Periodic Event Handler 				& No  			&  Single Processor\\ \hline
Level~1 & \parbox[t]{3cm}{Preemptive Priority\\ Scheduling} & \parbox[t]{3.5cm}{Aperiodic and One-Shot\\ Event Handlers}  & No 			&  Multi-Processor\\  \hline   
Level~2 & \parbox[t]{3cm}{Preemptive Priority\\ Scheduling} & \parbox[t]{3cm}{Mission Sequencers\\ and\\ Managed Threads} & Yes 			& \parbox[t]{2.5cm}{Multi-Processor\\ Global\\ Scheduling} \\    \hline
  \end{tabular}

\caption{Comparison of SCJ Compliance Level Features\label{ComplianceLevels}}
\end{table}

A Level 0 application's execution model is essentially a cyclic executive. In this model, only periodic handlers are supported; they are executed sequentially in a precise, clock-driven time line~\cite{locke1992software}. A single mission sequencer controls the sequential execution of one or more missions.

At SCJ Level~1, missions are controlled by a single mission sequencer. The available schedulable objects are periodic, aperiodic, and one-shot event handlers. At Level~1, schedulable objects are executed concurrently by a preemptive priority-based scheduler; any access to shared data has to be performed by \verb"synchronized" methods to avoid race conditions and to assure that the compiler generates code that forces that changes made to shared variables by one thread to propagate to all other threads that share access to those same variables. A notable restriction of the Level~1 programming model is that use of \verb"Object.wait()" and \verb"Object.notify()" is prohibited. Arbitrary use of such methods complicates the ability to perform schedulability analysis.

At Level~2, missions are executed sequentially by a top-level mission sequencer, as with Level~1. In addition, each mission may register nested mission sequencers during its initialisation phase. Once these nested mission sequencers begin running, they each execute a sequence of child missions, independently of the top-level mission sequencer. 
Computation in a Level~2 mission can be performed by periodic, aperiodic, and one-shot handlers, and no-heap managed real-time threads.
Each child mission has its own mission memory, distinct from its parent's mission memory.
A Level~2 application may use Java suspension features.

It is clear that those applications that can be scheduled using cyclic-executive techniques should be implemented at Level 0.
Furthermore, applications that can use simple analysable fixed-priority scheduling should use Level 1. Hence, the required scheduling techniques are a primary indicator of whether or not Level 0 should be used.
However, Level 2 also targets fixed-priority scheduling, so this cannot be used to decide between using Level 1 or Level 2.

To understand the purpose of Level 2, it is necessary to discover the generic application-level programming requirements for which Level 2 functionality is necessary.
In the current version of the specification, this is not provided in the rationale for the three compliance levels.

We broadly classify the additional functionality provided at Level 2 into three groups:~\begin{enumerate}
\item nested mission sequencers;
\item managed threads: including the use of the \verb"Object.wait()", \verb"Object.notify()", \\ \verb"HighResolutionTime.waitForObject()" and \verb"Services.delay()" methods; and,
\item global scheduling across multiple processors.
\end{enumerate}

\par\noindent
We explore the first group in Section~\ref{nested missions}, showing how they provide support for two example applications: a Space Shuttle, which has several modes of operation, with mode-specific schedulable objects and persistent schedulable objects running throughout all modes; and a Train Control system, which has multiple independent subsystems, each implemented using a nested mission sequencer. Programming several modes of operation is possible at Level~1, but combining this behaviour with tasks running throughout all modes (without restarting that task in each mode) is only possible at Level~2. Moreover, programming multiple subsystems is not possible at Level~1, due to  nested mission sequencers being unavailable.

In Section~\ref{threads} we focus on the second group of features above, examining the benefits of the \texttt{ManagedThread} class and presenting three motivating scenarios that show where they are useful: non-standard release profiles, suspension-based waiting, and encapsulation of local state.

The availability of global scheduling only at Level~2 reflects the fact that the state of the art in multiprocessor schedulability analysis is still advancing~\cite{davis2011survey}. Future safety-critical systems may be able to execute on multiprocessor platforms supported by new analysis techniques. We, however, do not address global scheduling in this paper.

We identify several areas where the SCJ specification needs modifications in order to fully support the programming requirements identified in Sections~\ref{nested missions}~and~\ref{threads}; these are summarised and expanded on in Section~\ref{issues}. The added functionality of Level 2 warrants a more formal description of the programming model and its required run-time support. In particular, the starting and termination of nested mission sequencers, and their associated missions, is much more complex than at Levels 0 and 1.
In Section \ref{FormalModellingOfSCJLevel2}, we present a formal model of the termination protocol and show that significant simplification to this aspect of the specification can be achieved with a simple change to the API. Related work is given in Section~\ref{related}, and we draw our conclusions in Section~\ref{conclusions}.

\section{Nested Mission Sequencers} \label{nested missions}

The ability to construct applications composed of nested mission sequencers is, perhaps, the most important aspect to be considered when choosing between Levels 0 or 1 and Level 2. In this section we identify two software architectural patterns that require the support of nested mission sequencers. We also sketch an example application for each of the patterns. We call these two patterns the \emph{Multiple-Mode Application Pattern} and the \emph{Independently Developed Subsystem Pattern}.

\subsection{The Multiple-Mode Application Pattern} \label{modechange}

\subsection*{Overview}

This pattern captures the typical architecture of systems that have to operate in multiple modes. Each mode consists of multiple persistent activities with well defined release frequencies and deadlines. In addition to these per-mode activities, there may also be persistent concurrent activities, which execute in all modes.
Well known schedulability analysis techniques can be used to guarantee the timing properties in the steady-state situations of execution in each mode. Analysis techniques also exist for handling the transitions between modes, but only on a single processor~\cite{tindell1992mode,real2004mode}.

\subsection*{Architecture Components}

\begin{figure}[h]
  \centering
  \includegraphics[scale=0.3]{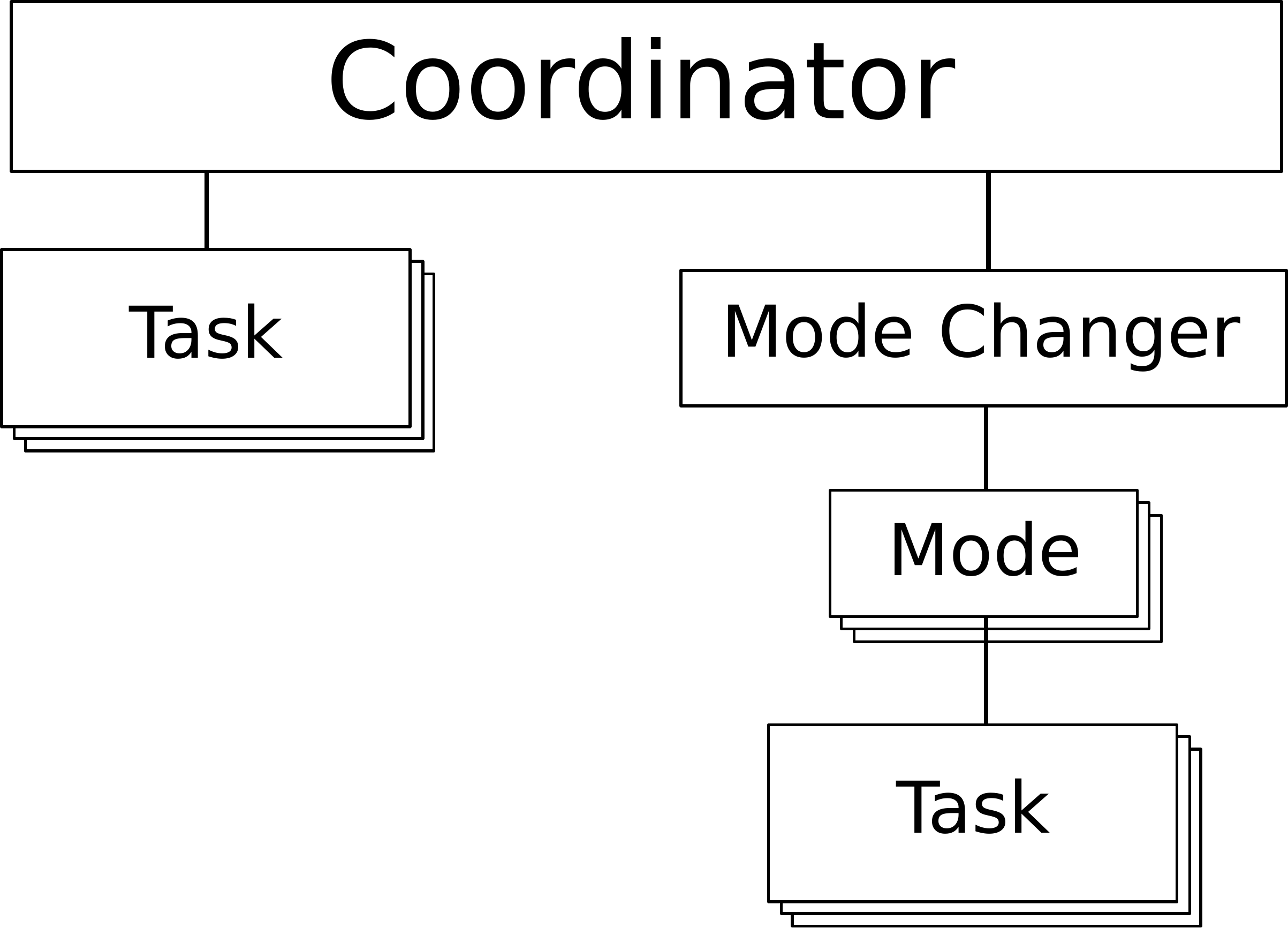}\\
  \caption{Multiple-Mode Operations Pattern}\label{modes_pattern}
\end{figure}

The components that characterise this pattern are shown in Figure~\ref{modes_pattern}. Tasks represent concurrent activities. A mode changer encapsulates several modes, and each mode encapsulates several mode-specific tasks. Only one mode per mode changer is active at any one time. There may also be persistent tasks, which are required to operate during all modes. The mode changer and any persistent tasks are controlled by a coordinator. Mode changes are typically requested by tasks from the currently active mode. 

In terms of SCJ, a mode changer can be conveniently implemented as a mission sequencer, and each mode as a mission. The tasks can be realised as SCJ schedulable objects. The coordinator component also has a natural correspondence with a mission, often the main mission, which registers the persistent tasks and the mode changer, and controls their operation.

\subsection*{Example Application}

\begin{figure*}[ht]
\centering
\includegraphics[scale=0.3]{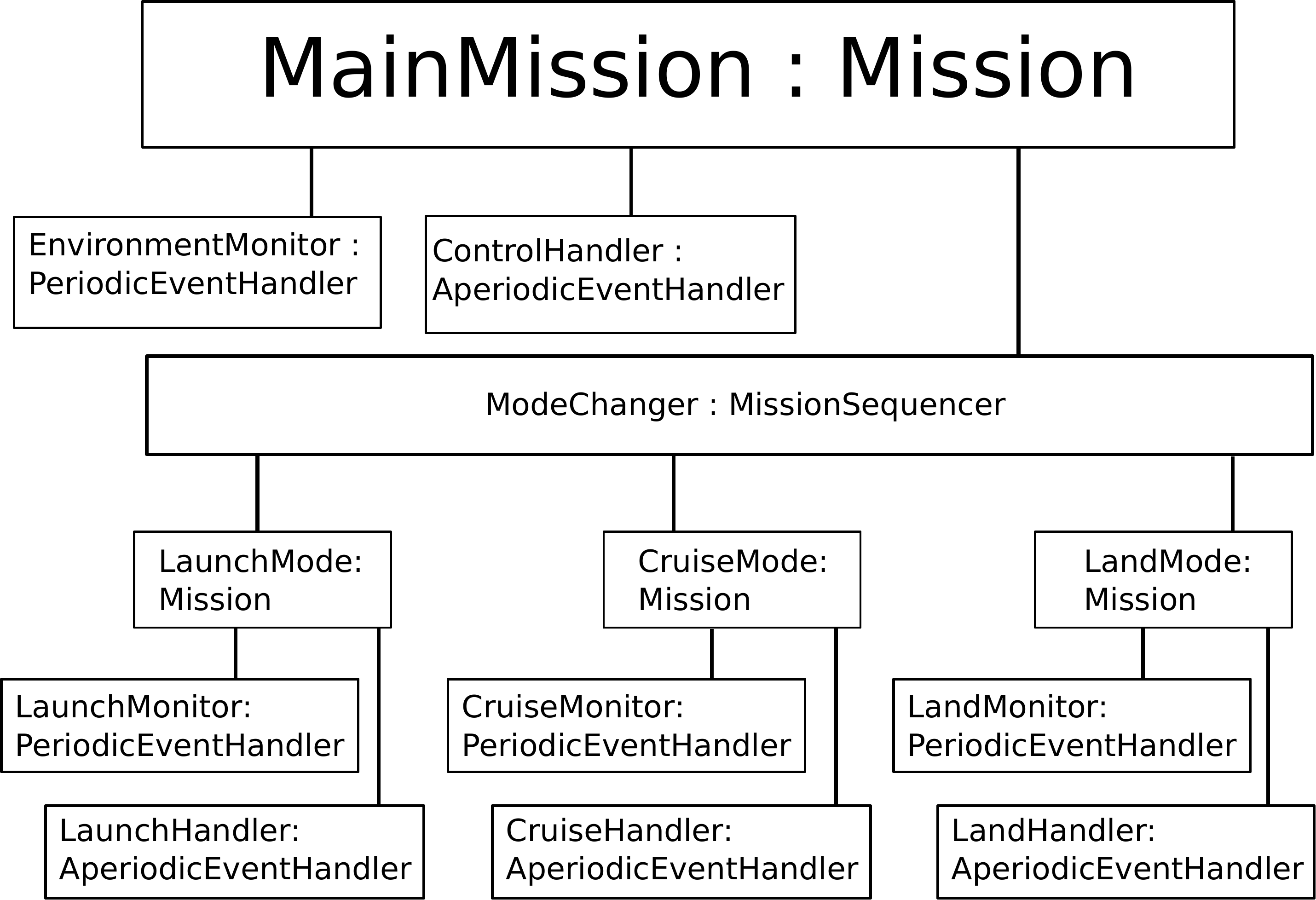}\\
\caption{Space Shuttle with Moded Operations}\label{shuttle}
\end{figure*}

An example application that uses this pattern is an idealised Space Shuttle\footnote{The code for this example can be found at \url{http://www.cs.york.ac.uk/circus/hijac/case.html} }, as illustrated in Figure \ref{shuttle}. It has three modes, each associated with a phase of its operation. Each mode has several schedulable objects that are only active during that mode -- only two are shown for each mode in Figure~\ref{shuttle}. In addition to the mode-specific schedulables, there are two persistent schedulables shown, \texttt{EnvironmentMonitor} and \texttt{ControlHandler}, which are active throughout all the modes.\\

\subsection*{Adequacy of SCJ Support}

Using missions to support individual modes of operation and mission sequencers to support the mode-change controllers has two main advantages. The first is that encapsulating each mode in a mission enhances the modularisation of the SCJ program and the traceability of its structure to its architectural model. This is important when each mode is a significant software component in its own right.

The second advantage is that SCJ supports a well-defined (if somewhat complicated -- see Section \ref{FormalModellingOfSCJLevel2}) process for mission termination, where schedulables can complete their current release before the mission completes. This is usually what is required when mode change requests are \emph{planned} events. (Planned mode changes occur at well defined points in a system's operation. In contrast, \emph{unplanned} mode changes usually occur as a result of error conditions being detected. Such errors may be anticipated, but the time of their occurrence can not be predicted. Hence the time at which a mode change is required cannot be predicted.)

On the other hand, in adopting the multiple-mode application pattern in the context of SCJ, there are issues of timing that need to be considered. First of all, in order to execute a new mode, it is necessary to create all the new objects (that are to reside in the mission memory) during the initialization phase of the mission (mode).  Hence, for unplanned mode changes or applications that require fast and predictable planned changes, there may be some efficiency or latency concerns.

In addition, there is no automatic single release time for all the schedulable objects in the application. The schedulable objects in the initial mode start at a different time from the persistent schedulable objects. To create a single start time, it is necessary to use absolute-time offsets for all periodic handlers.

For timing analysis, the mission sequencer, implementing the mode changer, must be viewed as an aperiodic activity whose minimum inter-arrival time is equal to the minimum time between mode change requests. Its deadline represents any time constraints on the mode change operation. As an SCJ mission sequencer is a managed event handler, it only has a priority; it does not have \emph{any} release parameters. These must be captured outside of the SCJ program and used in any schedulability analysis. We discuss this concern further in Section \ref{issues}.

Finally, it is not easy to provide the runtime scheduling needed to support compositional time analysis of the application, as SCJ does not support hierarchical scheduling. SCJ schedules persistent schedulable objects in competition with mode-specific schedulable objects. Hence, the whole application must be analysed in each mode along with each mode transition. We discuss this issue in Sections \ref{subsystem pattern} and \ref{4:servers} below.

\subsection{The Independently Developed Subsystem Pattern}\label{subsystem pattern}

\subsection*{Overview}
Assembling  systems that are composed of independently developed subsystems, each encapsulating related behaviours, is an important approach to developing systems that are more complex than those typically developed for Level~1. The ability to create nested mission sequencers at SCJ Level~2 is the key to supporting this approach to constructing systems.

\subsection*{Architecture Components}

\begin{figure}[t]
  \centering
  \includegraphics[scale=0.3]{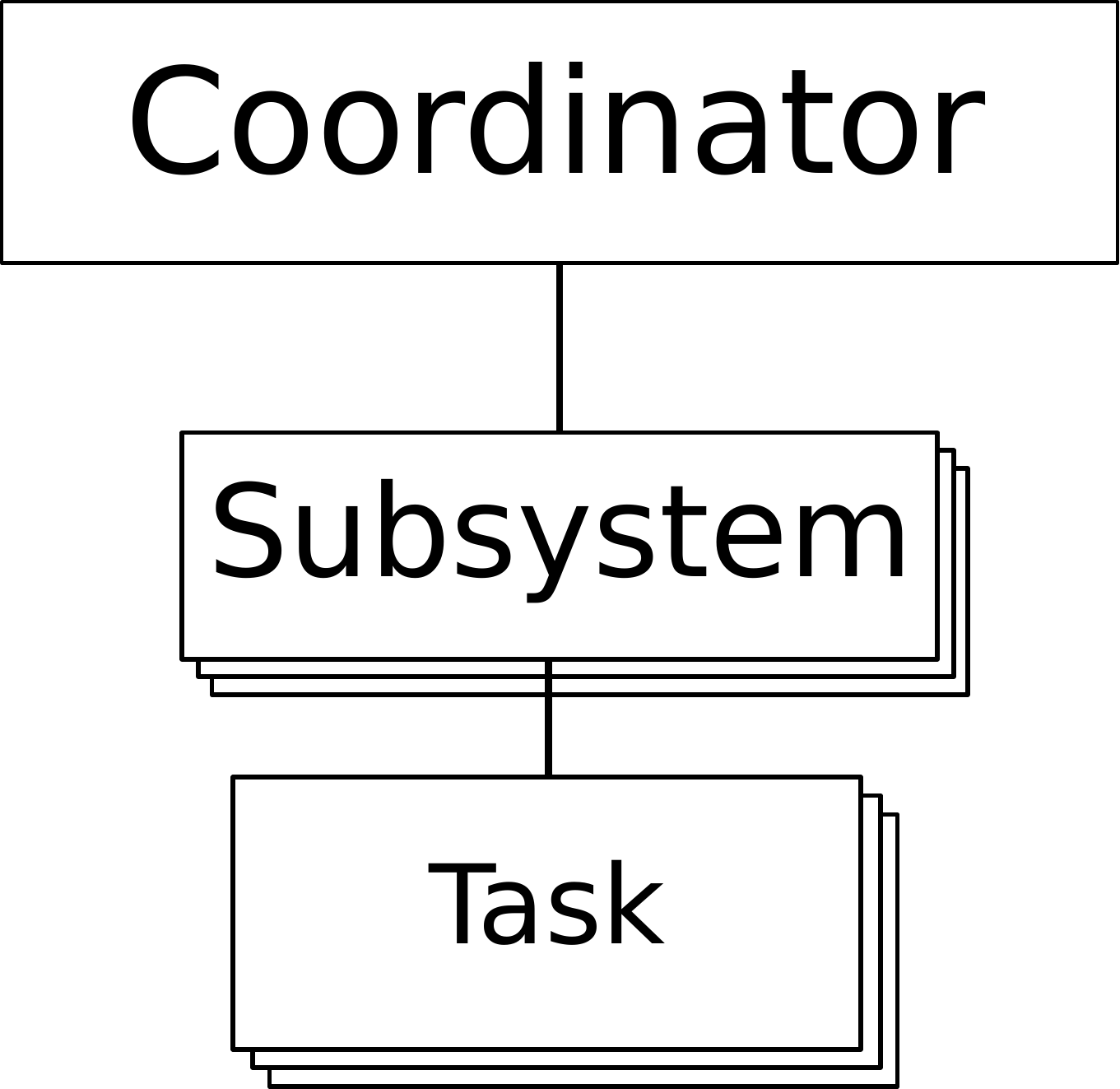}\\
  \caption{The Independently-Developed Subsystem}\label{subsystems}
\end{figure}

The software architecture that characterises this pattern is shown in Figure \ref{subsystems}. The subsystems are all controlled by a coordinator. Subsystems may contain other subsystems. Typically, each subsystem is, or can be, independently developed and contains several tasks that perform related behaviours. 

In terms of SCJ each subsystem can be decomposed into a mission sequencer and a single mission that manages the tasks within that subsystem. Each task can then be implemented by an appropriate managed schedulable: mission sequencer, thread, or handler. In this setting the coordinator component corresponds naturally to a mission, often the main mission, that registers the mission sequencers of each subsystem.

\subsection*{Example Application}

\begin{figure*}[bht]
\centering
\includegraphics[scale=0.3]{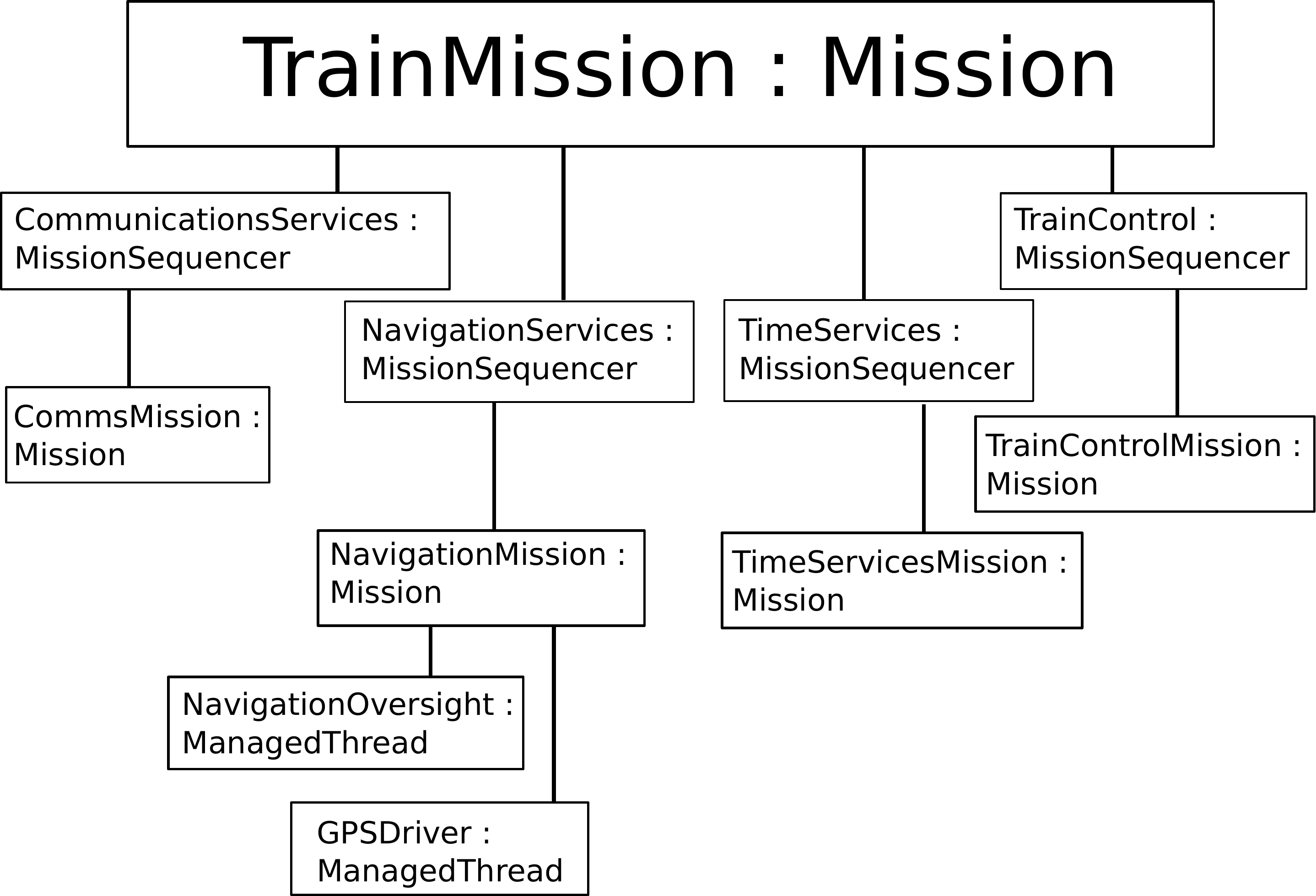}\\
\caption{Railway System with Multiple Subsystems}\label{railway}
\end{figure*}

A good example of this pattern is the railway system described by Hunt and Nilsen~\cite{hunt2012}:

\begin{quote}
``Collision avoidance in rail systems is a representative safety-critical
application.  A common approach to the challenge of avoiding train
system collisions divides all tracks into independently governed
segments.  A central rail traffic control system takes responsibility
for authorizing particular trains to occupy particular
rail segments at a given time.  Each train is individually responsible for
honouring the train segment authorizations that are granted to it.  Note
that rail segment control addresses multiple competing concerns. On the
one hand, there is a desire to optimize utilization of railway
resources. This argues for high speeds and unencumbered access. On the
other hand, there is a need to assure the safety of rail transport. This
motivates lower speeds, larger safety buffers between travelling trains,
and more conservative sharing of rail segments.''
\end{quote}
The example considers the structure of the on-board software (illustrated in Figure \ref{railway}), which supports the following
requirements:~
\begin{itemize}

\item maintain reliable and secure communication with the central
rail traffic control authority --- the \emph{CommunicationService} subsystem;

\item monitor progress of the train along its assigned route --- the \emph{NavigationService} subsystem;

\item control the train's speed in accordance with scheduled station stops,
rail segment authorizations, local speed limit considerations, and fuel
efficiency objectives --- the \emph{TrainController} subsystem;

\item infrastructure support for maintaining global time --- the \emph{TimeServices} subsystem.

\end{itemize}

\noindent
In the implementation described in~\cite{hunt2012}, each of these subsystems is realised as a nested mission sequencer registered to the main mission (\texttt{TrainMission}), and each subsystem controls a single mission that registers the subsystem-specific managed schedulables. There are multiple tiers of nested mission sequencers within the subsystems. Each tier represents further subsystems that can be developed independently. For brevity, Figure~\ref{railway} only shows two of the subsystem-specific managed schedulables and omits the deeper tiers of nested mission sequencers.

\subsection*{Adequacy of SCJ Support}

Although the encapsulation provided by missions is ideal for structuring subsystems,  as illustrated above there are issues that need to be addressed when adopting this approach. The first is that in order to compose a system from many subsystems (missions), each of these subsystems (missions) must be controlled by its own mission sequencer. This is natural if each mission has multiple modes of operation, but can become cumbersome otherwise.

The second issue has already been mentioned in Section \ref{modechange}. When a system is composed of subsystems, there is no automatic common release time for all the schedulable objects. If required, this has to be programmed explicitly. For multiple nested sequencers, this can become cumbersome because the system start time needs to be passed down to all schedulable objects in the system.

Whilst the above limitations can be seen as minor, the third limitation, which we discuss next, is more significant. Neither SCJ nor the RTSJ directly support hierarchical scheduling. Hence it is difficult to achieve decomposability of timing constraints when subsystems are independently developed.
The RTSJ does support the notion of processing groups, which allow several schedulable objects to share a CPU budget, but these are too general and difficult to use in a multiprocessor environment~ \cite{burns2003processing,wellings2008processing}.
Hierarchical scheduling techniques for single processor and partitioned multiprocessor systems are well established~\cite{davis2005hierarchical} and techniques are beginning to emerge for globally scheduled multiprocessor systems~\cite{burmyakov2012generalized,davis2011survey}. The lack of such a facility in SCJ severely limits its use in supporting the timing analysis of applications composed according to the independently-developed subsystem pattern. We return to this issue in Section~\ref{4:servers}.

The final issue with the functionality of SCJ, in relation to this programming pattern, is that the API supports a request to terminate a mission sequencer. The intention of this facility is to allow a schedulable object within a mission to request not only its mission to be terminated, but also the whole sequence of missions (of which it is part) to be terminated. The concern with the facility is that it can be misused by a schedulable to terminate an arbitrary mission sequencer. This complicates the semantics of the termination protocol needed to support mission termination and breaks the encapsulation of the mission concept. We return to this issue in Sections \ref{4:proposedChanges} and \ref{FormalModellingOfSCJLevel2}.

\section{Using Managed Threads} \label{threads}

There are several motivations for supporting managed threads in SCJ Level 2 applications. The first is to serve the needs of schedulable objects that do not have a standard release profile. The second is to allow suspension-based waiting for input and output operation completion. The final motivation is to allow more encapsulation of state information. We consider each of these, in turn, in this section.

\subsection{Non-Standard Release Profiles}

It is impossible to anticipate every possible scenario in which a schedulable object might need to be released. Here, we consider three common scenarios and discuss the difficulties of implementing them at Level 1. We choose one of these to illustrate how implementation using Level 2 is possible.

\subsection*{A Periodic Activity Released by an Event}

In SCJ, a periodic activity is either released immediately when it is started, or released after an absolute or relative delay from when it is started. There is no possibility of releasing a periodic activity via notification from another schedulable object 
or, indeed, an interrupt. However, it can be desirable for the initial release of a periodic activity to be triggered by a notification from another schedulable object, the absence of such a notification, or an interrupt. For example, the implementation of a task controlling a mechanical system that requires periodic updates but is started by an aperiodic button press can benefit from such a release pattern.

The above discussion suggests that the Level 1 support for periodic event handlers is not flexible enough to cope with anything other than simple time-released periodic activities. Simply introducing the \verb"Object.wait()" and \verb"Object.notify()" methods into Level 2, to allow a periodic handler to wait for a notification, is not sufficient. The problem is that deadline monitoring of event handlers starts from when the handler is first released. In SCJ, deadlines cannot be dynamically changed, so it is not possible to set an initial deadline and then change it after the notification has occurred.

The introduction of managed threads at Level~2, on the other hand, allows these more general release patterns to be addressed, as managed threads allow programmers to implement their own release mechanisms. We consider, for example, the periodic managed thread released by software notification illustrated in Figure \ref{periodicThread}, which shows an abstract extension of the \texttt{ManagedThread} class. The \texttt{firstRelease} method (lines 18-23) is called during the mission to indicate that the periodic activity should now start. The abstract \texttt{work} method declared on line 35 must be overridden to provide the functionality to be called each period. The \texttt{run} method (lines 38-47) is final and waits for the initial release before calling the \texttt{work} method periodically. This example illustrates the added flexibility that is available at Level 2; the periodic thread in Figure~\ref{periodicThread} can not be programmed at Level~1.

\begin{figure*}
\begin{Java}
public abstract class PeriodicThread extends ManagedThread {

  private AbsoluteTime nextRelease; //the next release time of this thread
  private AbsoluteTime nextDeadline; // the next deadline of this thread
  private final int period;
  private final int deadline;
  private DeadlineMissHandler deadlineMissDetection;
  private Mission myMission; //this thread's controlling mission
  private boolean hadFirstRelease = false;

  public synchronized void firstRelease() {
    hadFirstRelease = true;
    notify();
    nextRelease = Clock.getRealtimeClock().getTime(nextRelease);
    nextDeadline.set(nextRelease.getMilliseconds() + deadline);
    deadlineMissDetection.scheduleNextReleaseTime(nextDeadline);
  }

  private synchronized boolean waitFirstRelease() {
    while(!hadFirstRelease){
       try { wait(); }
       // or HighResolutionTime.waitForObject(this, timeout)
  	   // if a timeout is also required
       catch(InterruptedException ie) { // mission is to be terminated
       return false;
    	}
    }
    return true;
  }

  protected abstract void work();
    // override this to provide the function of the thread

  public final void run() {
    if (waitFirstRelease()) {
      while(!myMission.terminationPending()) {
        nextRelease.add(period,0);
        work();
        nextDeadline.add(period,0);
        deadlineMissDetection.scheduleNextReleaseTime(nextDeadline);
        Services.delay(nextRelease); // waitForNextPeriod
      }
    }
  }
}
\end{Java}

\caption{A Periodic Schedulable Object Released by Software Notification}\label{periodicThread}
\end{figure*}

Another example of a more complicated release pattern that can only be programmed in SCJ at Level~2 is the thruster control system described by Wellings~\cite[Page 235]{Wellings2004}. Here, an astronaut activates the thruster and supplies a duration for the engine ``burn''. The control of the engine itself requires a periodic activity to avoid the mechanical drift of valves. This requires an activity that is released by an event, executes periodically for a certain duration (determined either by time itself or by another event), and then waits to be started again. For the same reasons as those described above for an event-released periodic activity, the only way this release pattern can be supported in SCJ is with managed threads using \verb"Object.wait()"
and \verb"Object"\-\verb".notify()".

On the other hand, it should be noted that managed threads are a simplified version of the RTSJ's no-heap real-time thread, with the following restrictions: there is no automatic release mechanism (that is, no support for \verb"waitForNextPeriod") and there is no mechanism to add a deadline. Furthermore, in SCJ the per-release memory area is created when the thread starts and cleared when the thread terminates. Consequently, if needed, developers have to program their own support for more sophisticated memory management.

\subsection*{Consumers in a Producer-Consumer System}

Another common release pattern is where a producer schedulable object generates data that must be processed by a consumer schedulable object. Typically this data may come in bursts, and the consumer should process all the data as quickly as possible and block when there is no data available. These requirements cannot be met at Level 1, since it does not support a queue of outstanding release events for aperiodic event handlers. Level 2 allows this release pattern to be programmed using managed threads.

\subsection*{Background Activities: Run as Fast as You Can}

There are occasions where background activities are required to run as fast as possible. This is the case, for example, of a logging task that is required to process data from application logs whenever the scheduler allows it access to the processor. There is no notion of release events for these activities (other than their initial start). These activities can be programmed with Level~1 functionality using either an aperiodic event handler that is released only once, or a one-shot event handler with no start wait time. Both of these options, however, are a misuse of these mechanisms. Although there is no negative consequence for this misuse, a managed thread is a better abstraction to support this requirement.

\subsection{Suspension-based Waiting for IO where Busy-Waiting is Inappropriate}

In many systems, a device driver busy-waits for its associated device input (or output) to complete. This is because the expected delay is small and context switching away from the driver is considered inefficient. There are ways to integrate this delay into the scheduling of the driver (see \cite[Section  14.6]{burns2009real}), and allowing the driver to delay when it has no other
activity to perform may also be appropriate. On the other hand, when this delay is a relatively significant amount of time, it is necessary to allow the system to schedule some alternative activities. Since it is not possible to have a suspension-based delay at Level 1, this requirement can only be implemented at Level 2.

\subsection{Encapsulation of State Information}

Another characteristic that differentiates managed thre\-ads from event handlers is their use of memory. An event handler has its private memory area cleared at the end of each release, which means that state that must persist across releases must be saved in an outer memory area. A managed thread, however, only has its memory area cleared when it exits its \texttt{run()} method (that is, it terminates). This means that data can be stored locally and preserved over successive application-implemented `releases' of the thread.  

Of course, the effect of these two approaches is the same. The thread's memory area can last for as long as the memory area of its controlling mission, which is where persistent data used by an event handler is normally stored. However, this ability to encapsulate state is important from a software engineering perspective, since storing data that is private to a schedulable object in the mission memory of its controlling mission makes this data more widely visible than it should be.

As an example, we consider several schedulable objects that log their local state changes into local bounded buffers. When a buffer becomes full (which may take several releases of its associated schedulable object), the data is copied into a single global buffer in mission memory, which another schedulable object uses to write the system state changes to disk. If the logging schedulable objects are event handlers, the local buffers cannot be stored in their per-release memory areas, as such areas are cleared at the end of each of their releases. They need to be stored in the mission memory.  Using managed threads, the local buffers can be stored in the per-release memory areas, as these are not cleared until their associated managed threads terminate. In addition, the local buffers do not become exposed to access by other schedulables.

Application-implemented releases, such as those programmed in Figure \ref{periodicThread}, can also be augmented to use a nested private memory area for objects that can be cleared at the end of each application-implemented release. This is illustrated in Figure \ref{periodicThread2}, which just shows the augmented \texttt{run()} method (and an associated runnable) of lines 38-47 of Figure \ref{periodicThread}.

\begin{figure}
\begin{Java}
Runnable R = new Runnable () {
  public void run() { work(); }
};

public final void run() {
  if (waitFirstRelease()) {
    while(!myMission.terminationPending()) {
      nextRelease.add(period,100);
      ManagedMemory.enterPrivateMemory(privateMemorySize, R);
      nextDeadline.add(period,100);
      deadlineMissDetection.scheduleNextReleaseTime(nextDeadline);
      Services.delay(nextRelease);
    }
  }
}
\end{Java}
\caption{Augmented Periodic Schedulable Object}\label{periodicThread2}
\end{figure}

\section{Revisiting the SCJ Level 2 Support}\label{issues}

Sections \ref{nested missions} and \ref{threads} have explored some of the application requirements where the use of Level~2 functionality is desirable. Here, we review the issues identified as potential causes of problems, and explore changes that can be made to the SCJ specification to avoid these problems.

\subsection{Managed Thread Termination} \label{4:thread termination}

In SCJ, a managed thread terminates when it returns from its \texttt{run()} method. In Section~\ref{threads} we illustrate how this simple release pattern can be adapted, using Level~2 features, to provide more complicated release patterns. Figure~\ref{periodicThread} shows the example of a periodic thread that is first released by software notification using the \verb"Object.wait()" method (on line 18). With this approach, however, if our periodic thread is released and is waiting for its software notification when its controlling mission begins termination, then the thread may not finish its current release -- its \texttt{run()} method may remain active. More generally, this applies to any schedulable object that is using the suspension features available at Level~2; if they are waiting when their controlling mission begins termination, then their release may not finish.

SCJ~\cite{SCJDraft} defines the following activities to be performed on receipt of a mission termination request:

\begin{itemize}
\item invoke this mission's \texttt{terminationHook()} method;
\item invoke \texttt{signalTermination()} on each managed schedulable object that is registered for execution within this mission;
\item disable all periodic event handlers associated with this mission so that no further firings occur;
\item disable all aperiodic event handlers, so that no further firings are honoured;
\item clear the pending release event (if any) for each event handler so that the event handler can be effectively shut down following completion of any event handling that is currently active;
\item wait for all of the managed schedulable objects associated with this mission to terminate their execution;
\item invoke the \texttt{ManagedSchedulable.cleanUp()} method for each of the managed schedulable objects associated with this mission; and,
\item invoke the \texttt{cleanUp()} method associated with this mission.
\end{itemize}
\noindent
We note that this list does not require invoking the \texttt{interrupt()} methods of all the managed schedulables, which would cause all blocked managed schedulables to wake up with an exception and hence expedite termination. This has to be programmed by applications using the \texttt{Mission.terminationHook()} method, and can be inconvenient when the mission has many schedulable objects.

To aid the termination of managed threads and schedulables that are suspended when a termination request is received, we propose that either the SCJ infrastructure interrupts all schedulable objects associated with the mission or that all the schedulable  objects associated with the mission are informed of a pending termination request. The latter proposal can be achieved via a new method (\texttt{terminationSignalled()}), which each managed schedulable object must implement. The intention of this method is to allow the programmer to manually interrupt those schedulable objects that may be blocked when mission termination is signalled.

\subsection{Deadlines on Mission Sequencers} \label{4: deadlines MS}

As discussed in Section \ref{modechange} an SCJ mission sequencer does not have any release parameters. Therefore, it cannot have an associated deadline or deadline-miss handler. Systems that support multiple modes of operations often have deadlines associated with the mode changes. Hence, at Level 2 it is appropriate to allow some form of deadline-miss handler to execute if the mode change does not occur promptly.

Adding aperiodic release parameters to mission sequencers seems to undermine the mission programming model, particularly for sequencers that support a single non-terminating mission. Instead, what we propose is to add the methods shown in Figure~\ref{newMSMethods} to the \texttt{MissionSequencer} class. These methods identify deadline-miss handlers for mission termination and start. We note that, since mission changes can also occur in Level~1, these facilities might also prove useful in that context. 

\begin{figure}

\begin{Java}
/**
  * As for Mission.requestTermination
  *
  * In addition, the SCJ infrastructure will set a timer that will fire if mission
  * termination (including any cleanup) has not completed by the
  * deadline. On expiry of the timer, the infrastructure will release the aperiodic
  * event handler passed as a parameter.
  *
  * The timer will be cancelled if it has not fired when the mission terminates.
  */
  @SCJAllowed(Level_1)
  public final void requestTerminationOfCurrentMission(AbsoluteTime deadline,
       AperiodicEventHandler deadlineMiss);

/**
  * As for Mission.requestTermination
  *
  * In addition, the SCJ infrastructure will set a timer that will fire if next mission
  * has not started by the deadline. On expiry of the timer, the
  * infrastructure will release the aperiodic event handler passed as a parameter.
  *
  * The timer will be cancelled if it has not fired when the new mission starts.
  *
  * If there is no new mission, the timer is cancelled when the call to getNextMission
  * returns null.
  */
  @SCJAllowed(Level_1)
  public final void requestMissionChange(AbsoluteTime deadline,
       AperiodicEventHandler deadlineMiss);
\end{Java}
\caption{Proposed New Methods for the \texttt{MissionSequencer} Class \label{newMSMethods}}
\end{figure}

\subsection{Support for Compositional Timing Analysis} \label{servers}\label{4:servers}

Section~\ref{subsystem pattern} identifies a role for mission sequencers as a mechanism that can support the composition of safety-critical systems from independently-developed subsystems (or components). We represent a subsystem in SCJ with a mission sequencer controlling a single mission, which controls that subsystem's schedulable objects, as detailed in Section~\ref{subsystem pattern}. Hence, we consider the mission sequencer as the top of the subsystem. 

Hierarchical scheduling (and its associated schedulability analysis) is a well established technique that facilitates composition when components have real-time attributes (such as deadlines). Unfortunately, hierarchical scheduling is supported by neither SCJ nor the RTSJ. This is possibly because of the lack of support by real-time operating system vendors. We propose incorporating two elements of hierarchical scheduling into SCJ to improve its support for independently-developed subsystems and components: CPU budgets, to implement execution-time servers; and multi-level priorities, to isolate the scheduling of subsystems.

In the proposed approach, constructing a system made of subsystems can be achieved broadly in three steps. 
First, each subsystem is allocated an execution-time server, which is given a capacity, a priority order, and a replenishment period. These parameters need to be assigned carefully to obtain good schedulability~\cite{davis2008investigation}.
Next, the priority ordering of the schedulable objects in each subsystem must be determined. 
Finally, an integration step assigns concrete priorities to the schedulable objects based on their priority ordering and the priority order of their server.

The schedulable objects within a subsystem are only scheduled for execution (in priority order) when their execution-time server server would be scheduled at the top level (and has available capacity).
Once the parameters of the execution-time servers and the schedulables are set, the program needs to be analysed to determine schedulability at both the system and subsystem levels. We note that there is a relationship between the priorities of the execution-time server and of the subsystem's schedulable objects.

In the rest of this section we describe the integration of our proposal into SCJ. We consider only two tiers in the program hierarchy here, for brevity.

\subsubsection{CPU Budgets}~\\
\label{sub:bugets}
The first aspect of hierarchical scheduling we require is that each subsystem is allocated a budget, which is consumed whenever one of its schedulable objects is executing, and a period after which its budget is replenished. When a subsystem's budget has been totally consumed, all of its associated schedulable objects are suspended until its next replenishment occurs. In the RTSJ, this functionality can be supported by processing groups, if all the schedulable objects run on the same CPU. 

Implementations of the RTSJ that support processing groups ensure that members of a group, collectively, are not be given more CPU time per period than their group's budget. When supported, the RTSJ implements the \texttt{ProcessingGroupParameters} class, which is associated with each schedulable object in the processing group. This allows the RTSJ's schedulable objects to share a budget while retaining their individual priorities, deadlines, and periods.

Because processing groups support the requirements for compositional timing analysis, one possible solution is for SCJ to implement the following restricted version of the RTSJ \texttt{ProcessingGroupParameters} class, where the deadline of the processing group is equal to its replenishment period.

\begin{Java}
package javax.safetycritical;

public class ProcessingGroupParameters {
	public ProcessingGroupParameters (HighResolutionTime start, 
            RelativeTime replenishmentPeriod, RelativeTime budget){
            ...
	}	                               
	...
}
\end{Java}
\noindent

However, this technique inherits the limitation that the missions encapsulated within a mission~sequencer need to execute on the same processor.

\subsubsection{Simulating Multi-Level Priorities}~\\
\label{sub:priorities}
The second aspect of hierarchical scheduling that we require is multi-level priorities, which can be simulated in SCJ by manipulating the priorities of mission sequencers and schedulable objects. We propose:~
\begin{itemize}
\item using the priority of each mission sequencer to define a priority range: from the priority of this mission sequencer to the priority of the next highest priority mission sequencer, and;
\item transposing the priorities of all the schedulable objects in this subsystem into this range, while maintaining their original priority order, to ensure that they only run when their subsystem has the highest priority of all the subsystems.
\end{itemize} 
\noindent

This priority manipulation is performed statically, before the program is executed, in the integration step mentioned above. It may be the case that mission sequencer's priorities must be changed during integration to accommodate the schedulable objects. This is allowed as long as the priority order of the mission sequencers is maintained.

For example, we consider below a simple two-subsystem application using rate-monotonic scheduling. The parameters of the execution-time servers of each subsystem are shown in Table~\ref{tab:serverParams}. \\

\begin{figure}[h!]
\centering
\begin{tabular}{c||c|c}
			&	Period~(ms)	& Budget~(ms) 	\\
		\hline \hline
Server 1		&	100			&	40			\\
\hline	
Server 2		&	50			&	15			\\
\end{tabular}

\caption{Execution-Time Server Parameters \label{tab:serverParams}}
\end{figure}

~\\
\noindent
At the top level, the execution-time server the subsystems associated with $Server 1$ has a replenishment period of 100 milliseconds and a budget of 40 milliseconds.  The execution-time server of the subsystem associated with $Server 2$ has a replenishment period of 50 milliseconds and a budget of 15 milliseconds. The top-level is schedulable when the priority of $Server~2$ is greater than the priority of $Server~1$.

Now, we suppose that the subsystem associated with $Server~1$ contains three schedulable objects, $S1$, $S2$, and $S3$, which require a priority ordering where $S3$ has a higher priority than $S2$, which has a higher priority than $S1$. During system integration, the priorities of the servers and schedulables could be assigned so that the priority of $Server~2$ is greater than that of  $Server 1$ plus 3 in order to allow the priorities of the schedulable objects to be assigned between those of the two servers. An example of the priorities that can be assigned is shown in Table~\ref{tab:schedulablePriorities}.
\begin{figure}[h!]
\centering
\begin{tabular}{c||c|c}
			&	Schedulable	& Priority 	\\
		\hline \hline
Server 1		&				&	10			\\
\cline{2-3}
			&	S1			&	10			\\
			\cline{2-3}
			&	S2			&	11			\\
			\cline{2-3}
			&	S3			&	12			\\
\hline	\hline
Server 2		&				&	20			\\
\end{tabular}

\caption{Execution-Time Server and Schedulable Priorities \label{tab:schedulablePriorities}}
\end{figure}
~\\
\noindent
This concrete priority assignment simulates multi-level priorities, because the schedulables of the subsystem associated with $Server~1$ are not able to run if $Server~2$ is executing.

\subsubsection{Incorporation into SCJ}~\\

As detailed above, to support CPU budgets, SCJ needs to implement processing groups, and SCJ can already support multi-level priorities, by manipulating the priorities of an application's schedulable objects and mission sequencers. To aid integration of these two aspects of hierarchical scheduling into SCJ applications, a new subclass of mission sequencer can be added to encapsulate the concerns of a subsystem, as shown in the example below.
\begin{Java}
public class Subsystem extends MissionSequencer{
    public Subsystem (PriorityParameters pri, StorageParameters storage,
                ProcessingGroupParameters params, int priRange){
            ...
	}
	...
}
\end{Java}
\noindent
The constructor above takes a \texttt{ProcessingGroupParameters} object, as described in Section~\ref{sub:bugets}. To encapsulate the information needed for the priority manipulation, described in Section~\ref{sub:priorities}, the values of \texttt{pri} (which is the priority of this mission sequencer) and of \texttt{pri + priRange} define the prioirity range for schedulable objects encapsulated by this subsystem.

\subsection{Mission Sequencer Termination}\label{proposedChanges}\label{4:proposedChanges}

In Section \ref{subsystem pattern}, we argue that allowing arbitrary schedulables to request the termination of an arbitrary mission~sequencer violates the encapsulation supported by missions. We propose that schedulable objects are only allowed to request that their controlling mission is terminated. The mission itself then has the responsibility of deciding whether to request its sequencer to terminate. This gives a more structured approach to termination.

We propose removing the \texttt{requestSequencerTermination()} method, which allows a request to terminate a mission sequencer, to enforce this more structured termination policy. Instead, we recommend that the mission cleanup phase indicates whether its sequencer should continue with the next mission or terminate. For that, we propose that \texttt{Mission.cleanUp()} return a boolean value, which is passed to the mission sequencer to determine if the mission sequencer should continue or terminate.

We investigate the impact of this change in the next section. As the termination protocol is one of the more complex features of SCJ Level 2 programs, we consider formal models of the SCJ termination protocol for both the current specification and for the protocol that we propose here.

\section{Formalisation of Level 2}

\label{FormalModellingOfSCJLevel2}

In this section we present: a formal model of the current termination protocol as presented in the SCJ draft specification~\cite{SCJDraft}, in Section~\ref{current-termination}; a formal model of the termination protocol incorporating our proposed changes, in Section~\ref{new-termination}; and a comparison of these two models, in Section~\ref{compare}. Our models are written in the state-rich process algebra \Circus{}, for which a model of SCJ Level~1 already exists~\cite{zeyda2013circus}. In Section~\ref{circus} we give a brief overview of the \Circus{} notation.

With these models we show that the current termination protocol is more complicated than necessary. Indeed, it was the process of formally modelling SCJ Level~2 that first illuminated the complexities of the mission sequencer termination protocol. These complications only become apparent at Level~2 because of its capacity to nest mission sequencers arbitrarily deeply, which means that mission sequencers can be terminated by schedulables both above and below themselves in the program's hierarchy at any point during the execution phase.

\subsection{\Circus{} Introduction}
\label{circus}
\Circus{}~\cite{cavalcanti2003refinement} combines elements from CSP~\cite{hoare1978csp}, Z~\cite{spivey1992z}, and a refinement calculus~\cite{morgan1990PS} to allow modelling of both state and patterns of interaction. Figure~\ref{CircusSyntax} sketches the BNF description of the syntax of \Circus{}. Below, we describe the elements of the syntax pertinent to the discussion of our formal model. A comprehensive account of \Circus{} can be found in~\cite{Oliveira2007-qg}.

\Circus{} programs, defined in Figure~\ref{CircusSyntax} by the syntactic category $\mathsf{Program}$, are formed by a sequence of \Circus{} paragraphs. Each \Circus{} paragraph, defined in Figure~\ref{CircusSyntax} as elements of $\mathsf{CircusPar}$, may be either a Z paragraph (the $\mathsf{Par}$ category), a channel declaration, a channel set declaration, or a process declaration. The syntactic category $\mathsf{N}$ contains the valid Z (and \Circus{}) identifiers.

\Circus{} programs use bi-directional channels to allow their processes to communicate; we discuss the different types of communications later in the section. All of the channels used in a \Circus{} program must be declared. Channel declarations are defined by the $\mathsf{CDecl}$ syntactic category in Figure~\ref{CircusSyntax}. Here, $\mathsf{Exp}$ is the category of Z expressions. If a channel takes any parameters, their types must be declared. For convenience, channels my be collected into a channel set -- defined by the $\mathsf{CSExp}$ category. Channel sets allow easy specification of the channels used to interact with a process.

Each \Circus{} process has a name and a body ($\mathsf{\circprocess\ N \circdef ProcDef}$) and may take parameters ($\mathsf{ Decl @ ProcDef }$). In our model, this is used where, for example, the process modelling a mission or a mission sequencer takes a parameter representing its unique identifier. Hence, for example, $\circprocess MissionFW \circdef mission : MissionID \circspot ~ \ldots$ declares the mission process with a parameter $mission$ of type $MissionID$.

The body of a Circus process ($\mathsf{\circbegin\ PPar^* \circstate\ SchemaExp\ PPar^* @ Action} \circend$) is delimited by the $\circbegin$ and $\circend$ keywords; it may contain a state, which is modelled using a Z schema; and some actions, modelled using a free combination of Z state operations, constructs of a simple imperative language, and CSP constructs ($\mathsf{PPar^*}$). While Z schemas can be used to define data operations over the state of a \Circus{} process using predicates, assignments to variables can also be made directly ($\mathsf{N^+} := \mathsf{Exp^+}$, from the $\mathsf{Command}$ category in Figure~\ref{CircusSyntax}). 

A \Circus{} process always has a main action at the end of the process after a $@$, that dictates the combination of Z schemas and CSP actions that define the behaviour of the process; these actions may reference other local actions for the purpose of structure. Both the state and actions of a \Circus{} process are local to that process. This makes \Circus{} processes similar to classes in object-oriented programming, where a class has some local variables and methods.

\begin{figure}
  \begin{syntax}
      %%%%%%%%%%%%%%%%%%%%%%%%%%%%%%%%%%%%%%%%%%%%%%%%%%%%%%%%%%%%%%%%%%%%%%
          \mathsf{Program} & \quad \mathsf{::=} \quad &
          \mathsf{CircusPar^*}
          \smallskip \\ %
      %%%%%%%%%%%%%%%%%%%%%%%%%%%%%%%%%%%%%%%%%%%%%%%%%%%%%%%%%%%%%%%%%%%%%%
          \mathsf{CircusPar} & \quad \mathsf{::=} \quad &
          \mathsf{Par} \ \mathsf{|} \ \circchannel\ \mathsf{CDecl} \
          \mathsf{|} \
          \circchannelset\ \mathsf{N} \circdef \mathsf{CSExp} \ \mathsf{|} \ \mathsf{ProcDecl}
          \smallskip \\ %
      %%%%%%%%%%%%%%%%%%%%%%%%%%%%%%%%%%%%%%%%%%%%%%%%%%%%%%%%%%%%%%%%%%%%%%
          \mathsf{CDecl} & \mathsf{::=} &
          \mathsf{SimpleCDecl}\ \mathsf{|} \
          \mathsf{SimpleCDecl};\mathsf{CDecl}
          \\ %
          \mathsf{SimpleCDecl} & \mathsf{::=} & \mathsf{N^+} \
          \mathsf{|} \ \mathsf{N^+: Exp} \ \mathsf{|} \
          \mathsf{[N^+]N^+:Exp} \ \mathsf{|} \
         ~\ldots% \mathsf{SchemaExp}
         
          \\ %
          \mathsf{CSExp} & %
          \mathsf{::=} & \mathsf{ \lchanset ~ \rchanset } %
          \ \mathsf{|} \ \mathsf{ \lchanset N^+ \rchanset} %
          \ \mathsf{|} \ \mathsf{N}
          \ \mathsf{|} \ \mathsf{CSExp \cup CSExp} %
          \ \mathsf{|} \ \mathsf{CSExp \cap CSExp}
          \\ %
          & \mathsf{|} & \mathsf{CSExp \setminus CSExp}
          \smallskip \\ %
     %%%%%%%%%%%%%%%%%%%%%%%%%%%%%%%%%%%%%%%%%%%%%%%%%%%%%%%%%%%%%%%%%%%%%%
          \mathsf{ProcDecl} & \quad \mathsf{::=} \quad & \mathsf{\circprocess\ N \circdef ProcDef} \ \mathsf{|} \ \ldots% \mathsf{\circprocess\ N[N^+] \circdef ProcDef} %
          \\ %
      %%%%%%%%%%%%%%%%%%%%%%%%%%%%%%%%%%%%%%%%%%%%%%%%%%%%%%%%%%%%%%%%%%%%%%
          \mathsf{ProcDef} & %
          \mathsf{::=} & \mathsf{ Decl @ ProcDef } %
    %  \ \mathsf{|} \ \mathsf{ Decl \odot ProcDef } %
          \ \mathsf{|} \ \mathsf{ Proc }   ~\ldots%
          \\ %
     %%%%%%%%%%%%%%%%%%%%%%%%%%%%%%%%%%%%%%%%%%%%%%%%%%%%%%%%%%%%%%%%%%%%%%
          \mathsf{Proc} & \mathsf{::=} & \mathsf{\circbegin\ PPar^*\
          \circstate\ SchemaExp\ PPar^* @ Action\ \circend} %
          \\ %
          ~\ldots%
         % & \mathsf{|} & \mathsf{Proc; Proc}
        %  \ \mathsf{|} \ \mathsf{Proc} \extchoice \mathsf{Proc}
       %   \ \mathsf{|} \ \mathsf{Proc} \intchoice \mathsf{Proc}
        %  \\ %
        %  & \mathsf{|} & \mathsf{Proc} \lpar \mathsf{CSExp} \rpar \mathsf{Proc} %
       %   \ \mathsf{|} \ \mathsf{Proc} \interleave \mathsf{Proc} %
       %   \ \mathsf{|} \ \mathsf{Proc} \hide \mathsf{CSExp}
       %   \\ %
       %   & \mathsf{|} & ( \mathsf{Decl @ ProcDef} ) ( \mathsf{Exp^+})%
       %   \ \mathsf{|} \ \mathsf{N} ( \mathsf{Exp^+} )
       %   \ \mathsf{|} \ \mathsf{N} %
       %   \\ %
       %   & \mathsf{|} & ( \mathsf{Decl \odot ProcDef} ) \lfloor \mathsf{Exp^+} \rfloor %
       %   \ \mathsf{|} \ \mathsf{N} \lfloor \mathsf{Exp^+} \rfloor %
       %   \ \mathsf{|} \ \mathsf{Proc} [ \mathsf{N^+} := \mathsf{N^+} ]
        %  \ \mathsf{|} \ \mathsf{N} [ \mathsf{Exp^+} ]
       %   \\ %
        %  & \mathsf{|} & \ \mathsf{Decl} @ \mathsf{Proc} %
       %   \ \mathsf{|} \ \Extchoice \mathsf{Decl} @ \mathsf{Proc} %
        %  \ \mathsf{|} \ \Intchoice \mathsf{Decl} @ \mathsf{Proc}
       %   \\ %
      %    & \mathsf{|} & \lpar \mathsf{CSExp} \rpar \mathsf{Decl} @ \mathsf{Proc} %
       %   \ \mathsf{|} \ \Interleave \mathsf{Decl} @ \mathsf{Proc} %
          \smallskip \\ %
      %%%%%%%%%%%%%%%%%%%%%%%%%%%%%%%%%%%%%%%%%%%%%%%%%%%%%%%%%%%%%%%%%%%%%%
          \mathsf{NSExp} & %
          \mathsf{::=} & \mathsf{ \{ ~ \} } %
          \ \mathsf{|} \ \mathsf{ \{ N^+ \} } %
          \ \mathsf{|} \ \mathsf{N}
          \ \mathsf{|} \ \mathsf{NSExp \cup NSExp} %
          \ \mathsf{|} \ \mathsf{NSExp \cap NSExp}
          \\ %
          & \mathsf{|} & \mathsf{NSExp \setminus NSExp}
          \smallskip \\ %
      %%%%%%%%%%%%%%%%%%%%%%%%%%%%%%%%%%%%%%%%%%%%%%%%%%%%%%%%%%%%%%%%%%%%%%
          \mathsf{PPar} & \mathsf{::=} & \mathsf{Par}
          \ \mathsf{|} \ \mathsf{N \circdef ParAction}
          \ \mathsf{|} \ \mathsf{\circnameset\ N \circdef NSExp}
          \smallskip \\ %
      %%%%%%%%%%%%%%%%%%%%%%%%%%%%%%%%%%%%%%%%%%%%%%%%%%%%%%%%%%%%%%%%%%%%%%
          \mathsf{ParAction} & \mathsf{::=} & \mathsf{Action}
          \ \mathsf{|} \ \mathsf{Decl @ ParAction}
          \smallskip \\ %
      %%%%%%%%%%%%%%%%%%%%%%%%%%%%%%%%%%%%%%%%%%%%%%%%%%%%%%%%%%%%%%%%%%%%%%
          \mathsf{Action} & \mathsf{::=} & %\mathsf{SchemaExp}
          \ \mathsf{|} \ \mathsf{Command}
          \ \mathsf{|} \ \mathsf{N}
          \ \mathsf{|} \ \mathsf{CSPAction}
         ~\ldots% \ \mathsf{|} \ \mathsf{Action}~[ \mathsf{N^+} := \mathsf{N^+} ]
          \\ %
      %%%%%%%%%%%%%%%%%%%%%%%%%%%%%%%%%%%%%%%%%%%%%%%%%%%%%%%%%%%%%%%%%%%%%%
          \mathsf{CSPAction} & \quad \mathsf{::=} \quad & %Skip \
        %  \mathsf{|}
          \ Stop \ \mathsf{|} \ Chaos
         % \ \mathsf{|} \ \mathsf{Comm} \then \mathsf{Action} %
          \ \mathsf{|} \ \mathsf{Pred}\ \&\ \mathsf{Action}
         % \\ %
         % & \mathsf{|} & \mathsf{Action}; \mathsf{Action} %
          %\ \mathsf{|} \ \mathsf{Action} \extchoice \mathsf{Action} %
          \ \mathsf{|} \ \mathsf{Action} \intchoice \mathsf{Action}
          \\ %
     
       %   & \mathsf{|} & \mathsf{Action} \lpar \mathsf{NSExp} | \mathsf{CSExp} | \mathsf{NSExp} \rpar \mathsf{Action} %
    %      \\ %
        %  & \mathsf{|} & \mathsf{Action} \linter \mathsf{NSExp} |
        %  \mathsf{NSExp} \rinter \mathsf{Action}
   %       \\ %
        %  & 
       &   \mathsf{|} & \mathsf{Action} \hide \mathsf{CSExp} %
        %  \ \mathsf{|} \ \mathsf{ParAction} (\mathsf{Exp^+})
        %  \ \mathsf{|} \ \mu \mathsf{N} @ \mathsf{Action}
     %     \\ %
%          &
           \mathsf{|}  \Semi \mathsf{Decl} @ \mathsf{Action} %
        %  \ \mathsf{|} \ \Extchoice \mathsf{Decl} @ \mathsf{Action} %
        %  \ \mathsf{|} \ \Intchoice \mathsf{Decl} @ \mathsf{Action} \\%
        %  & \mathsf{|} & \lpar \mathsf{CSExp} \rpar \mathsf{Decl} @ \lpar \mathsf{NSExp} \rpar \mathsf{Action} %
         % \ \mathsf{|} \ \Interleave \mathsf{Decl} @ \linter \mathsf{NSExp} \rinter \mathsf{Action}%
         ~\ldots
          \smallskip \\ %
      %%%%%%%%%%%%%%%%%%%%%%%%%%%%%%%%%%%%%%%%%%%%%%%%%%%%%%%%%%%%%%%%%%%%%%
          \mathsf{Comm} & \quad \mathsf{::=} \quad & \mathsf{N} \ \mathsf{CParameter^*} %
         ~\ldots% \ \mathsf{|} \ \mathsf{N} \ ~[\mathsf{Exp}^+] ~\mathsf{CParameter^*} \\ %
          \smallskip \\ %
          \mathsf{CParameter} & \quad \mathsf{::=} \quad & ?\mathsf{N} \ %
          \ \mathsf{|} \ ?\mathsf{N}:\mathsf{Pred} %
          \ \mathsf{|} \ !\mathsf{Exp} %
          \ \mathsf{|} \ \mathsf{.Exp} \\ %
          \smallskip \\ %
      %%%%%%%%%%%%%%%%%%%%%%%%%%%%%%%%%%%%%%%%%%%%%%%%%%%%%%%%%%%%%%%%%%%%%%
          \mathsf{Command} & \quad \mathsf{::=} \quad & \mathsf{N^+} := \mathsf{Exp^+} %
          \ \mathsf{|} \ \circif\ \mathsf{GActions}\ \circfi %
          \ \mathsf{|} \ \circvar\ \mathsf{Decl} @ \mathsf{Action}  \\%
        %  & \mathsf{|} & \mathsf{N^+} : [~ \mathsf{Pred} , \mathsf{Pred} ~] %
        %  \ \mathsf{|} \ \{\mathsf{Pred}\}%
       %   \ \mathsf{|} \ [\mathsf{Pred}] \\%
          & \mathsf{|} & \circval\ \mathsf{Decl} @ \mathsf{Action} %
         % \ \mathsf{|} \ \circres\ \mathsf{Decl} @ \mathsf{Action} %
         ~\ldots% \ \mathsf{|} \ \circvres\ \mathsf{Decl} @ \mathsf{Action} %
          \smallskip \\ %
      %%%%%%%%%%%%%%%%%%%%%%%%%%%%%%%%%%%%%%%%%%%%%%%%%%%%%%%%%%%%%%%%%%%%%%
          \mathsf{GActions} & %
          \mathsf{::=} & \mathsf{Pred} \then \mathsf{Action} %
          \ \mathsf{|} \ \mathsf{Pred} \then \mathsf{Action} \extchoice
          \mathsf{GActions}
      %%%%%%%%%%%%%%%%%%%%%%%%%%%%%%%%%%%%%%%%%%%%%%%%%%%%%%%%%%%%%%%%%%%%%%
  \end{syntax}
  \caption{\label{CircusSyntax}Partial BNF Syntax of \Circus}
\end{figure}

%Actions 
CSP has many operators that are adopted in \Circus{}, which all belong to the syntactic category $\mathsf{CSPAction}$ in Figure~\ref{CircusSyntax}. Table~\ref{circusOperators} provides a description of the operators in this category that we use in our model, some of which are omitted in Figure~\ref{CircusSyntax}. We describe these in more detail below to support the following discussion of our model. 

\begin{table}[h]
\centering
  \begin{tabular}{ l | l | l  }
    \hline
Action			& Syntax 							& Description \\ \hline \hline 
Skip			& $\Skip$ 								& A simple operator that terminates \\ \hline
Simple Prefix 	& $c \then A$ 						& Simple synchronisation with no data\\ \hline
Prefix			& $c.x \then A$ 						& Synchronisation with some data $x$ \\ \hline
Input Prefix 	& $c?x \then A$						& Synchronisation with a value bound to $x$ \\ \hline
Output Prefix	& $c!x \then A$ 						& Synchronisation outputting the value of $x$\\ \hline
External Choice & $A \extchoice B$ 					& Offers a choice between two actions $A$ and $B$\\ \hline
Sequence		& $A \circseq B$  						& Executes $A$ then $B$ in sequence\\ \hline
Parallelism 	& $A \lpar ns_a | cs | ns_b \rpar B$ 	& Parallelism, synchronising on the channels in $c$\\ \hline
Interleaving 	& $A \lpar ns_a | ns_b \rpar B$ 		& Parallelism with no synchronisation between \\ \hline
Iterated Interleaving & $\Interleave x : S @ A(x) $ 	& Interleaving of all actions $A(x)$ where $x \in S$\\ \hline
  \end{tabular}  

\caption{Syntax of \Circus{} operators derived from CSP \label{circusOperators}}
\end{table}

A simple operator is $\Skip$, which terminates and does nothing else. A prefix $c \then A$ waits for a communication on the channel $c$ and then proceeds to behave like the action $A$. If a channel has a parameter then this must be provided. The parameter can be included as either an input ($c?x \then A$), an output ($c!x \then A$), or added to the channel name to indicate a specific communication on that channel ($c.x \then A$). This latter form is often used in our models to restrict an action to synchronise on a channel only if it is parametrised by the identifier of the \Circus{} process to which it belongs. A related operator is sequential composition $\circseq$, which connects any two processes, instead of just a channel communication and a process like the prefix operator $\then$. Hence $A \circseq B$ executes the action $A$ until it terminates and then continues on to execute $B$. 

The external choice operator $\extchoice$ allows an action to offer its environment the choice of two different channel communications. Hence $c_1 \then A ~ \extchoice ~ c_2 \then B$ proceeds to $A$ if there is a communication on $c_1$ or $B$ if there is a communication on $c_2$. \Circus{} also contains a simple conditional statement as shown in the definition of the syntactic category $ \mathsf{Command}$ in Figure~\ref{CircusSyntax}. It takes a familiar if\ldots then\ldots else form. Hence $\circif~(x = TRUE) \circthen A~\circelse~(x = FALSE) \circthen B~ \circfi$ performs the action $A$ if $x = TRUE$ and the action $B$ if $x = FALSE$.

Two actions $A$ and $B$ may be placed in parallel: $A \lpar ns_a | cs | ns_b \rpar B$, specifies a synchronisation set of channels $cs$ over which they both have to agree to communicate, and name sets describing the variables that each side of the parallelism may alter that must be disjoint to avoid write conflicts. Hence, in the execution of $A \lpar \emptyset | \lchanset a, b \rchanset | \emptyset \rpar B$, $A$ and $B$ perform their actions in parallel with each other, but they must both agree to communicate on the channels $a$ and $b$ at the same time; further, the use of the empty set~($\emptyset$) indicates that neither $A$ nor $B$ can alter any variables.

\subsubsection{SCJ in \Circus{}}~\\
\label{scjInCircus}

A formal model of SCJ Level~1 has been produced \cite{zeyda2013circus} to allow the translation of arbitrary SCJ programs into \Circus{} in order to facilitate analysis. The \Circus{} models are composed of a model of the infrastructure classes of SCJ -- the `Framework Model' -- which remains the same and is reused for each translation, and a model of the code provided by the application -- the `Application Model' -- which changes for each new program translated. The Framework Model encapsulates the unchanging aspects of \textit{any} SCJ program, whereas the Application Model is generated afresh each time to model a \textit{specific} SCJ Level~1 program.

A \Circus{} model of SCJ Level~2 is in development, based on the approach taken by the Level~1 model; therefore, it has a Framework and an Application component.

The Level~2 model has \Circus{} processes for each of the main infrastructure classes in SCJ, and each object in the program is represented by its own instance of the relevant \Circus{} process. Generally speaking, each \Circus{} process retains the name of the SCJ class it models, suffixed with~``$FW$'' for framework processes and ``$App$'' for application processes.

The methods of each of the objects that we model are represented by \Circus{} actions. Ordinarily, a call to a method is modelled by two channels: a channel modelling the call to the method, suffixed by $Call$; and a channel modelling the return from the method, suffixed by $Ret$. For example, mission sequencers need to know that their current mission's \texttt{initialize()} method has completed before continuing. The action modelling the \texttt{initialize()} method, therefore, starts with a synchronisation on $initializeCall$ and terminates with a synchronisation on $initializeRet$. However if, in our model, the caller of the method does not require that the method returns before it continues, then a call to that method is modelled by a single channel. 

\begin{figure}

\centering
\includegraphics[width=\textwidth]{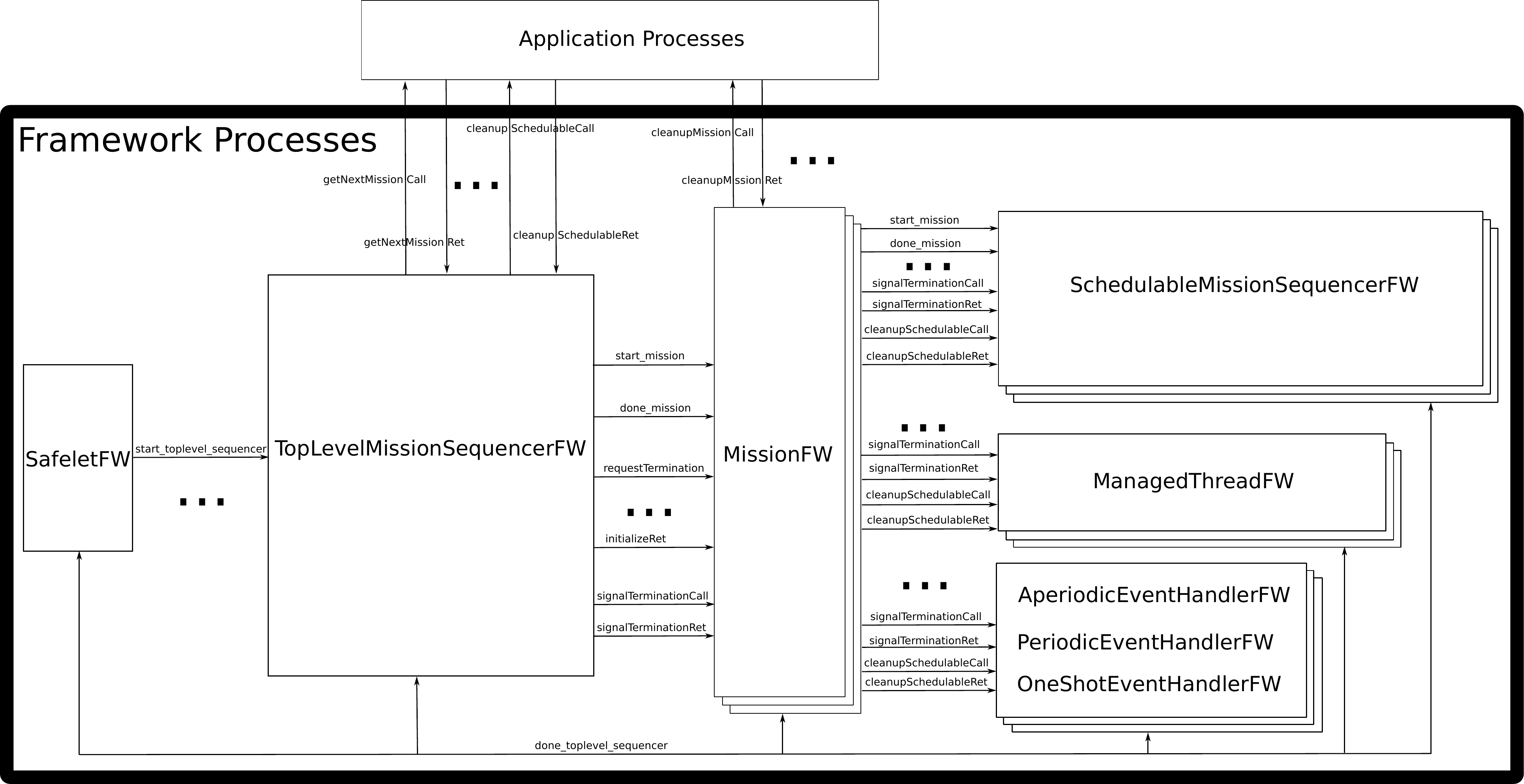}
\caption{Level 2 Model Structure}
\label{Level2Structure}
\end{figure}

Figure~\ref{Level2Structure} shows the processes that model the SCJ infrastructure interacting within the Framework Model and some of the communications between the Framework and Application models. For brevity, omitted communication channels are represented by ellipses, the three types of event handlers are represented by the single component at the bottom right of the figure, and all application processes are represented by the single component shown at the top of the figure. Because a mission sequencer can be used in two contexts at Level~2 -- both as a mission sequencer at the top of the program hierarchy and as a schedulable object nested inside a mission -- this class is modelled by two processes: one for the top-level mission sequencer, $TopLevelMissionSequencerFW$; and one for a schedulable mission sequencer, $SchedulableMissionSequencerFW$. This simplifies both processes because they each only have to be involved in communications relevant to their context. The $MissionFW$ and the processes representing the schedulable objects may have multiple instances in one model. Each of the SCJ methods that we model is represented by a CSP action in the relevant \Circus{} process.

\subsection{Model of the Current Termination Protocol}
\label{current-termination}

The current termination protocol requires very complex models of the mission sequencer process. This holds for both the $TopLevelMissionSequencerFW$ and the $SchedulableMissionSequencerFW$. Complexity arises because mission sequencers may be terminated at arbitrary times during their execution. In this section we describe our model of the protocol and explain the source of its complexity.\\

\subsubsection{Top Level Mission Sequencer}~\\
\label{sssec:tlms}

\noindent
The $TopLevelMissionSequencerFW$ process has one parameter, $sequencer$, which is the identifier of this mission sequencer process, and two state components: $currentMission$, which holds the identifier of the mission this sequencer is currently executing; and $terminating$, which is a boolean value that records if this mission sequencer has been asked to terminate. The $GetNextMission$ action models the \texttt{getNextMission()} method and is shown below. 
\begin{circusaction}
GetNextMission ~ \circdef ~ \\
  \t1 getNextMissionCall~.~sequencer \then \\
  \t1 getNextMissionRet~.~sequencer~?~next \then\\
  \t1 currentMission := next \circseq \\
  \t1 StartMission \circseq \\
  \t1 \circif ~ terminating ~ = ~ FALSE ~ \circthen \\
  \t2 GetNextMission \\
  \t1 \circelse ~ terminating ~ = ~ TRUE ~ \circthen \\
  \t2 \Skip  
\end{circusaction} 
It communicates with the application model using the channels $getNextMissionCall$ and $getNextMissionRet$ to get the identifier of the next mission that this mission sequencer should execute. This identifier is stored in the variable $currentMission$. Then the $StartMission$ action is called; it uses the $start\_mission$ channel to start the current mission. This action is defined as follows.
\begin{circusaction}
StartMission ~ \circdef \\
\t1 \circif ~ currentMission ~ \neq ~ nullMissionId ~ \circthen \\
\t2 \left( \begin{array}{l}
	start\_mission~.~currentMission \then\\
	initializeRet~.~currentMission \then \\
	\left( \begin{array}{l}
		RequestSequenceTermination\\
		\t1 \lpar \{terminating\}| \lchanset end\_termination \rchanset| \emptyset \rpar \\
		\left( \begin{array}{l}
		done\_mission~.~currentMission \then \\
		end\_termination~.~sequencer \then 
		\Skip	
		\end{array} \right)
	\end{array} \right)
\end{array} \right)\\
\t1 \circelse ~ currentMission ~ = ~ nullMissionId ~ \circthen \\
\t2 terminating := TRUE \circseq 
\Skip\\
\t1 \circfi \\
\end{circusaction}
\noindent
Once the current mission enters its execution phase (indicated by the communication on the $initializeRet$ channel) the $RequestSequenceTermination$ action is offered in parallel with a communication on the $done\_mission$ channel, which is used by the current mission to indicate that it has terminated. 

The parallelism here specifies that $RequestSequenceTermination$ and the communications in the brackets below the parallel operator synchronise on $end\_termination$, that $RequestSequenceTermination$ alters the $terminating$ variable, and the behaviours below the parallel operator do not alter any variables. %The parallel operator is explained in more detail in Table~\ref{circus}.

Once a communication on $done\_mission$ occurs, $StartMission$ waits for the action $RequestSequenceTermination$ to be ready to engage in $end\_termination$; this ends both sides of the parallelism and control returns to the $GetNextMission$ action. $GetNextMission$ then checks the value of the variable $terminating$, which is set 

in the $RequestSequenceTermination$ action shown in Section~\ref{sssec:rst}, to determine whether it should recurse or exit.

\subsubsection{Schedulable Mission Sequencer}~\\
\label{ssec:sms}

\noindent
The $SchedulableMissionSequencerFW$ process represents a mission sequencer nested within a mission; it is slightly more complicated than the $TopLevelMissionSequencerFW$. It has a $sequencer$ parameter and a $currentMission$ state component, like the top-level mission sequencer process. Instead of $terminating$ it has two state components: $terminatingAbove$, which indicates if this mission sequencer's controlling mission has asked it to terminate, and $terminatingBelow$, which indicates if this mission sequencer's current mission (or one of its managed schedulables) has asked it to terminate. Two variables are required because we have to treat the requests for termination differently, depending on their source.  With two variables,  we can model the different protocols separately. 

Here, the $GetNextMission$ action behaves identically to that used in the process $TopLevelMissionSequencerFW$  (Section~\ref{sssec:tlms}) aside from the conditional statement below, which checks both $terminatingAbove$ and $terminatingBelow$ to handle the possibility of the nested mission sequencer being asked to terminate from above or below itself in the program hierarchy.
\begin{circusaction}
\circif ~ terminatingAbove ~ = ~ FALSE ~ \land ~ terminatingBelow ~ = ~ FALSE ~ \circthen \\
  \t1 GetNextMission \\
\circelse ~ terminatingAbove ~ = ~ TRUE ~ \lor ~ terminatingBelow ~ = ~ TRUE ~ \circthen \\
  \t1 \Skip  
\end{circusaction}
The $StartMission$ action of the $SchedulableMissionSequencerFW$, shown below, contains a parallelism of three actions that offer the choice of waiting for its controlling mission to signal its termination (handled by the $SignalTermination$ action), the $RequestSequenceTermination$ action (which we discuss in Section \ref{sssec:rst}), and waiting for its current mission to communicate its termination on $done\_mission$.
\begin{circusaction}
StartMission ~ \circdef \\
 \circif currentMission \neq nullMissionId \circthen \\
\quad \left( \begin{array}{l}
	start\_mission~.~currentMission \then \\
	initializeRet~.~currentMission \then \\
	\left( \begin{array}{l}
		\left( \begin{array}{l}
			SignalTermination\\
			 \lpar \{terminatingAbove\} | \lchanset end\_terminations \rchanset | terminatingBelow \rpar 	\\		
		 RequestSequenceTermination\\
		 \end{array} \right)\\
           \lpar \{terminatingAbove, terminatingBelow\} |\lchanset end\_terminations \rchanset| \emptyset \rpar\\
        \left( \begin{array}{l}
        done\_mission~.~currentMission \then\\
        end\_terminations~.~sequencer \then \\
        \Skip\\		
        \end{array} \right)
	\end{array} \right)
\end{array} \right)\\
\circelse currentMission = nullMissionId \circthen\\
\t1 terminating := TRUE\\
\circfi
\end{circusaction}

The $SignalTermination$ action below handles the interaction with the controlling mission of this nested mission sequencer when it indicates that this mission sequencer should terminate. 
\begin{circusaction}
SignalTermination ~ \circdef \\
\t1 	\left( \begin{array}{l}
  	\left( \begin{array}{l} end\_terminations~.~sequencer \then \Skip\end{array} \right)	\\	
  	\extchoice\\
  	\left( \begin{array}{l}
  		signalTerminationCall~.~sequencer \then\\
  		terminatingAbove := TRUE \circseq \\
  		requestTermination~.~currentMission \then\\
  		signalTerminationRet~.~sequencer \then\\
  		\Skip
  	\end{array} \right)\\ \circseq 
  	  end\_terminations~.~sequencer \then \\ \Skip	\\
\end{array} \right)\\ \\
\end{circusaction}

The $SignalTermination$ action handles the nested mission sequencer being terminated from above and the $done\_mission$ communication handles the nested mission sequencer's current mission telling it to terminate from below. The $RequestSequenceTermination$ action handles the nested mission sequencer being told to terminate its sequence of missions by a managed schedule. We discuss this action next.\\

\subsubsection{Request Sequence Termination}~\\
\label{sssec:rst}

\noindent
The $RequestSequenceTermination$ action, shown below, waits for a communication on the $requestSequenceTermination$ channel. After this, the value of $terminating$ is set to $TRUE$ and the mission is queried to see if it is active and has not been asked to terminate already -- using the channels $terminationPending$ and $missionActive$. If these conditions are met, the action communicates on $requestTermination$, which tells the current mission to begin terminating. Then $RequestSequenceTermination$ recurses, so that subsequent calls to \texttt{requestSequenceTermination()} in the SCJ application can be handled, and so that the action can be terminated using $end\_termination$.
\begin{circusaction}
RequestSequenceTermination ~ \circdef \\
\left( \begin{array}{l}
	\left( \begin{array}{l}
		requestSequenceTermination~.~sequencer \then \\
		terminating := TRUE \circseq \\
		terminationPending~.~currentMission~?~missionTerminating \then \\
		missionActive~.~currentMission~?~missionIsActive \then \\
		\left( \begin{array}{l}
			\circif ~ missionTerminating ~ = ~ FALSE ~ \land ~ missionIsActive ~ = ~ TRUE  ~\circthen \\
			\quad requestTermination~.~currentMission \then \\\t1 \Skip \\
			\circelse ~ missionTerminating ~ = ~ TRUE ~ \lor ~ missionIsActive ~ = ~FALSE ~ \circthen\\
			\quad \Skip\\	
			\circfi		
		\end{array}\right)\\ \circseq 
		RequestSequenceTermination
	\end{array} \right)\\
	\extchoice\\
	\left( \begin{array}{l}end\_termination~.~sequencer \then  \Skip\end{array} \right)\\
\end{array} \right) \\
\end{circusaction}
\noindent
In the $SchedulableMissionSequencerFW$ process, the $RequestSequenceTermination$ action differs only in that, where $terminating$ is set to $TRUE$, the variable $terminatingBelow$ is altered instead. This is to handle the schedulable mission sequencer being terminated from a schedulable that is above it in the program hierarchy using $SignalTermination$, or below it, using $RequestSequenceTermination$. This can be seen in the excerpts presented in Section~\ref{ssec:sms} where $SignalTermination$ sets $terminatingAbove$ and $GetNextMission$ checks both of these variables.

\subsubsection{Clean Up}~\\
\label{ssec:oldcleanup}

\noindent
Our model of a mission uses three actions to model its three phases of operation: initialisation, execution, and clean up. As soon as one phase ends, the mission transitions to the next phase. Hence, the mission's $Cleanup$ action begins directly after its $Execute$ action has finished. First, the $CleanupSchedulables$ action is called, which iterates over the set $schedulables$ and executes the \texttt{cleanUp()} application method using synchronisations on the $cleanupSchedulableCall$ channel followed by the $cleanupSchedulableRet$ channel for each schedulable using its identifier $s$ as a parameter. The interleave operator ($\interleave$) is used to interleave all of the clean up phases.

\begin{circusaction}
CleanupSchedulables ~ \circdef ~ \\
\t1	\Interleave s : schedulables \circspot\\
\t2		cleanupSchedulableCall~.~s \then\\
\t2		cleanupSchedulableRet~.~s \then \Skip
\end{circusaction}
\noindent

Once the clean up of each managed schedulable registered to this mission has completed, the $Cleanup$ action executes the \texttt{cleanUp()} method of the mission itself using the $cleanupMissionCall$ and $cleanupMissionRet$ channels. Afterwards, the $Finish$ action is executed; it informs the mission's application process to terminate ($end\_mission\_app$) and then uses $done\_mission$ to inform the mission's controlling mission sequencer that is has finished.

\begin{circusaction}
Finish ~ \circdef ~ \\
\t1    end\_mission\_app~.~mission \then\\
\t1    done\_mission~.~mission~ \then \Skip 
\end{circusaction}

\noindent
This model captures the termination protocol as it currently stands. While the model is tractable, we argue that the same functionality can be achieved with a simpler termination protocol. In Section \ref{new-termination} we describe our model of the termination protocol including our proposed changes.

\subsection{Model of Proposed Changes to Termination Protocol}
\label{new-termination}

This section describes a new model for the SCJ termination protocol incorporating our proposed changes. As explained in Section \ref{proposedChanges}, we propose the removal of the  \texttt{requestSequenceTermination()} method to prevent mission sequencers from being terminated by arbitrary schedulables. To enforce an organised termination of mission sequencers we propose that the \texttt{mission.cleanUp()} method return a boolean value which is passed to the mission sequencer to determine if the mission sequencer should continue or terminate.

In adapting our model to our proposed protocol, the state of both flavours of mission sequencer process have been altered. In the $TopLevelMissionSequencerFW$ process, $terminating$ has been replaced with $continue$. In the $SchedulableMissionSequencerFW$ process, both the $terminatingAbove$ and $terminatingBelow$ variables have been replaced with $continueAbove$ and $continueBelow$. These variables indicate to the sequencer that it should continue executing its sequence of missions (if they are both $TRUE$).

If $continue$ is $FALSE$, or either $continueAbove$ or $continueBelow$ is $FALSE$ in the case of the $SchedulableMissionSequencerFW$, then the mission sequencer does not execute any more missions. In the $SchedulableMissionSequencerFW$ the variable $ContinueBelow$ holds the return value from the current mission that is communicated to the mission sequencer at the end of the cleanup phase on the $done\_mission$ channel -- in the $TopLevelMissionSequencerFW$ process this value is held in the $continue$ variable. The $SchedulableMissionSequencerFW$'s $continueAbove$ variable is only changed during the $SignalTermination$ action, which handles the mission sequencer's controlling mission requesting it to terminate.

Removing the $RequestSequenceTermination$ action is a clear simplification of the model; the $requestSequenceTermination$ channel is no longer needed and is removed from the model entirely. Besides this, the actions (in the model of the current termination protocol) that use the $RequestSequenceTermination$ action are also simplified in our new model. We give more details of these simplifications in the following three sections. 

\subsubsection{Top Level Mission Sequencer}~\\

\noindent
The $StartMission$ action in the $TopLevelMissionSequencerFW$ process is simplified in comparison to the previous version in Section~\ref{sssec:tlms}, as can be seen from the excerpt presented below. 
\begin{circusaction}
StartMission ~ \circdef ~\\  
\t1   \circif ~ currentMission != nullMissionId ~ \circthen\\
\t2   \left( \begin{array}{l}
     start\_mission~.~currentMission \then\\
     done\_mission~.~currentMission~?~continueReturn \then\\
     continue := continueReturn \circseq 
    \Skip\\
   \end{array} \right)\\
\t1   \circelse ~ currentMission = nullMissionId ~ \circthen\\
    \t2 continue := FALSE \circseq \Skip\\
\t1   \circfi
\end{circusaction}
This action simply starts the current mission using the $start\_mission$ channel and then waits for it to terminate and communicate on the $done\_mission$ channel. 

\subsubsection{Schedulable Mission Sequencer}~\\

\noindent
The $StartMission$ action in the $SchedulableMissionSequencerFW$ process (which models a nested mission sequencer) is shown below. It is necessarily more complex than that of the $TopLevelMissionSequencerFW$ process, but still simpler than the previous version in Section~\ref{ssec:sms}.

\begin{circusaction} 
StartMission ~ \circdef ~\\  
\t1   \circif ~ currentMission != nullMissionId ~ \circthen\\

 \t2 \left( \begin{array}{l} 
   
     start\_mission~.~currentMission \then\\
     initializeRet~.~currentMission \then \\
     
     \left( \begin{array}{l} 
     
     	SignalTermination\\
     	\t1 \lpar \emptyset | \lchanset end\_terminations \rchanset | \{continueBelow\} \rpar\\
     	\left( \begin{array}{l}   
	    done\_mission~.~currentMission~?~continueReturn \then\\
	    continueBelow := continueReturn \circseq \\
	    end\_terminations \then 
	     \Skip \end{array} \right) \\
   	  \end{array} \right)
  \end{array} \right)\\
\t1   \circelse ~ currentMission = nullMissionId ~ \circthen\\
    \t2 continueBelow := FALSE \circseq\\
    \t2 \Skip\\
\t1  \circfi    
\end{circusaction}

\noindent
After the mission has been initialised (indicated by the $initializeRet$ channel) this action proceeds to a parallelism that offers $SignalTermination$ to handle this mission sequencer's controlling mission being terminated and a communication on $done\_mission$ that indicates that the mission sequencer's current mission has terminated.

\subsubsection{Clean Up}~\\
\label{sssec:cleanup}

\noindent
To model \texttt{Mission.cleanUp()}, which now returns a boolean value, the $MissionFW$ process's $cleanupMissionRet$ channel takes a boolean parameter. 
\begin{circusaction}
Cleanup ~ \circdef ~ \\
%\t1     deregister!schedulables \then\\
\t1     CleanupSchedulables \circseq\\
\t1     cleanupMissionCall~.~mission \then\\
\t1     cleanupMissionRet~.~mission~?~continueSequencer \then\\
\t1     Finish(continueSequencer) 
\end{circusaction}
This value is communicated to the $MissionSequencer$ process via the $done\_mission$ channel. This channel is the means of communication that allows a mission to inform its controlling mission sequencer of its completion, and, as revised, also communicates continuation information.
\begin{circusaction}
Finish ~ \circdef ~ \\
\t1    end\_mission\_app~.~mission \then\\
\t1    done\_mission~.~mission~?~continueSequencer  \then\\
\t1    \Skip
\end{circusaction}
When the $MissionSequencer$ receives the boolean value from $done\_mission$, it stores it in the variable $continue$, which is checked by the $GetNextMission$ action after the $StartMission$ action finishes. This variable is used to decide whether the $MissionSequencer$ should continue its execution and get another mission or terminate. This minor addition to the model presents little extra complexity, while supporting our proposal to simplify the termination protocol significantly.

Section \ref{compare} compares the two termination protocols in more detail, using our formal models.

\subsection{Comparison of Termination Protocols}
\label{compare}

The current termination protocol allows any schedulable object to call the method \texttt{requestSequenceTermination()} (the \Circus{} action for which is presented in Sect.~\ref{sssec:rst}) of any mission sequencer in the program, regardless of its place in the hierarchy. The mission sequencer that receives this call informs its current mission to terminate. This is captured in the excerpt in Sect.~\ref{sssec:rst} by the communication on the $requestTermination$ channel, which indicates to the mission that is should terminate. The mission, once it is instructed to terminate, passes this on to its schedulables -- at least one of which may have called the \texttt{requestSequenceTermination()} method of the mission sequencer in the first place. This creates a needless cycle of termination requests.

In the new termination protocol that we propose, the instigation of termination still begins in a schedulable object, but this request is only passed up one tier at a time. For example, if a reason to terminate the application is detected by a schedulable object, this is passed to its controlling mission -- by setting some flag in the mission for example. Once it has terminated, the mission communicates this request for termination to the mission sequencer controlling it, during the mission's clean up phase. This is captured in our models by the communication on the $done\_mission$ channel of the boolean parameter $continueSequencer$ to the $MissionSequencer$ process that controls the mission. In this way, the request for termination passes up the program hierarchy, with each tier terminating before the next tier begins handling its termination.

This prevents the situation present in the current protocol in which a schedulable object that initially discovers the need for termination is requested to terminate later when the termination request cascades back down from the mission sequencer it had called \texttt{requestSequenceTermination()} on initially. Our new approach does create a small amount of programmer overhead, since the programmer must ensure that schedulable objects can inform their controlling mission that it should return \texttt{false} from its \texttt{cleanUp()} method. A simple way to remove this small overhead is for the default implementation of \texttt{Mission.cleanUp()} to return \texttt{false}.

We note that even in the new termination protocol, the schedulable object that discovers the need for termination triggers the termination of its controlling mission and then is asked to terminate itself. To avoid this, the schedulable objects can be programmed to check for the termination of their controlling mission periodically and begin to shut themselves down; this is in fact the only way to terminate a \texttt{ManagedThread}. Another solution is to have the schedulable that discovers the need for termination to terminate itself after it has triggered the termination of its controlling mission.

Our changes have a subtle effect on the termination order of the objects in the program. As an example, we consider a program with two nested subsystems. With the current protocol, a schedulable object within one of them may call the method \texttt{requestSequenceTermination()} on the top-level mission sequencer and begin a cascade of termination requests that leads to the nested sub-systems terminating in parallel. In the same situation, using the new termination protocol, the termination requests must pass up the hierarchy from the schedulable object that initiates termination to the top-level mission sequencer. This means that the subsystem that contains the schedulable object that requested termination has to terminate before the request for termination passes to the top-level mission and the termination of the other subsystem -- and any schedulable objects started by the top-level mission -- begins.

In summary, the \texttt{requestSequenceTermination()} method complicates the SCJ termination protocol by allowing arbitrary termination of mission sequencers. Our models, while tractable, are complex when modelling this feature of the language at Level~2. With our proposed changes incorporated, our models become much simpler and are easier to analyse. Our proposed changes to the SCJ termination protocol represent a positive simplification of the language while retaining the ability to terminate a mission sequencer from the application.

In order to show how far our proposed changes simplify the model of SCJ, we have constructed two specifications based on a simple example program. This program contains a single mission, controlling two managed threads that share a one-place buffer in the mission's memory. One specification uses the model of the current framework and the other specification uses the model of the new framework, with our proposed changes. Both of these specifications have been translated to CSPm, the machine-readable version of CSP, in order to utilise the Failures Divergences Refinement checker (FDR)~\cite{fdr3} to model check the specifications. Because CSP does not have a notion of state in the same way that \Circus{} does, the CSP versions of our models use state processes to model the reading of and assignments to variables that \Circus{} allows, which means that the CSP models have more states than the \Circus{} versions.

The results obtained are from running a check for divergence-freedom while hiding any channels relating to the state processes. The model of the current framework shows 4,539,021 states, whereas the model of the new framework shows only 249,869 states. Our proposed changes decrease the number of states in the model by 94.5\% in comparison to the original model. Such a decrease in the number of states in our model shows a simplification that is useful, both for further modelling efforts and for programmer understanding of the SCJ paradigm.

\section{Related Work}\label{related}

There have been previous efforts to provide safe language subsets for safety-critical systems, similar to the intent of Safety-Critical Java (SCJ). MISRA C~\cite{MISRA} is a restricted subset of standard C that originated in the automotive industry, but now now provides guidelines for the use of C in other critical systems.MISRA C has gained wide popularity in aircraft, medical systems, and other critical software domains~\cite{Hatton2004}.

Several subsets of Ada have been developed since the language was first defined. One of the most widely used is SPARK Ada, which highly restricts the amount of language features available to the programmer. The intent is to reduce the risk of failures resulting from errors in programs. This is balanced by ensuring that the language has the right level of abstraction to provide the expressive power needed to hide the details of implementations. SPARK also acknowledges the desire for safety-critical programs to be verifiable and restricts the language with this objective~\cite{barnes_high_2003}. SPARK has become one of the most popular choices for high-integrity real-time systems.

The Ravenscar~\cite{dobbing_ravenscar_1998} profile is another subset of Ada. It has a similar level of complexity to SCJ Level~1. It aims to aid program reliability -- defined as predictable and consistent functioning. The control flow of a program is divided into two phases:~initialisation and execution. All concurrent entities are allocated in the initialisation phase and they are started at the beginning of the execution phase. The concurrent entities in a Ravenscar program may only be periodic or sporadic; aperiodic entities are not supported. These concurrent entities are scheduled by  a pre-emptive priority-based scheduler.

Drawing on the restrictions of the Ravenscar profile, the Ravenscar-Java profile~\cite{kwon_phd_2006} was created to improve the reliability of Java-based systems using the Real-Time Specification for Java (RTSJ). The RTSJ is the basis for SCJ Programs written in the Ravenscar-Java profile conform to the RTSJ standard, with extra restrictions to ensure the program adheres to the Ravenscar rules. Other Java profiles have been proposed: for example that by Schoeberl et al.~\cite{DBLP:conf/isorc/SchoeberlSTR07}, who also considers the possibility of missions as application modes of operations~ \cite{Schoeberl:2007:MMS:1778978.1778991}.

As far as we are aware, other than the work by Hunt and Nilsen~\cite{hunt2012}, there has been no previous work that has considered how components should be implemented cleanly in SCJ. There have been several approaches suggested for the RTSJ -- see~\cite{plsek2012} as an example and for a review of related approaches. Most of these projects either focus on the functional aspects of component declaration and system composition, or they focus on the use of the RTSJ's memory areas.

There have been attempts to integrate the OSGi Java-based framework~\cite{OSGi} with the RTSJ, but again little attention has been given to the composition of timing constraints. The notable exception is the work by Richardson and Wellings~\cite{richardson2012}, which considers real-time admission control of components within a Real-Time OSGi framework. They recognise the limitation of the RTSJ's processing groups. To achieve the same effect as hierarchical scheduling of execution-time servers
they use a combination of processing groups, priority scaling and periodic timers. Essentially each server's priority is represented by a range of the RTSJ's priorities. A component allocated to a server must use this range when assigning priorities to its schedulable objects.
The cost overrun handler that can be assigned to a processing group changes the priorities of its associated schedulable objects to a background priority.
A separate periodic event handler is created whose release coincides with the replenishment period. This resets the schedulable objects to
their original priorities. Effectively, this approach can be used to implement a sporadic-server. It forms the basis of the approach that we have proposed in Section \ref{servers} for SCJ.

Other formalisations of languages for safety-critical systems exist. Blazy and Leroy~\cite{blazy2009mechanized} present a formal semantics of the C subset, Clight. Ellison and Rosu~\cite{ellison2012executable} present an executable formal semantics of C, which allows model checking of the translated C program. Tews et al.~\cite{tews2008formal} provide a formal semantics of a subset of C++. This semantics is embedded in the Prototype Verification System (PVS) prover and a prototype translator tool allows the translation of program code to the PVS semantics. Automatic translation from safety-critical programming languages is part of our agenda for future work.

The memory model of SCJ is based upon that of the RTSJ. Engel has produced a formalisation of a restricted version of the RTSJ memory model~\cite{engel2008deductive, engel2009deductive} to prove the absence of runtime errors caused by misuse of the memory model. This formalisation is implemented with the KeY theorem prover~\cite{ahrendt2005key}. The restricted memory model Engel considers is similar to, but much less restrictive than, that used in SCJ. This indicates that Engel's approach may be capable of being adapted to model the SCJ memory model. Our model does not cover the memory model of SCJ, which is formalised in~\cite{CWW13}.

Brooke et al~\cite{brooke2007csp} use CSP -- one of the components of \Circus{} -- to model the semantics of the real-time extension to Eiffel, SCOOP. This follows a similar direction to our work, but since SCJ is a more restricted language, the CSP model in~\cite{brooke2007csp} is more complex due to the generality it supports.

Kalibera et al~\cite{kalibera2010exhaustive} present a technique to allow the model checking of SCJ Level~0~or~1 programs. They extend the Java PathFinder tool with a scheduling algorithm that allows it to explore the possible schedulings of an SCJ program. Their approach is concerned with scheduling errors, assertion failures, and scheduling dependant memory access errors. While that work actually focusses on SCJ, our approach is more general and allows the checking of a wider range of program properties.\\

\section{Conclusions}\label{conclusions}

SCJ Level 2 has received little public scrutiny. Most papers address SCJ either Level~0~or~1.
Whilst it is clear from the SCJ language specification what constitutes a Level~2 application (in terms of its use of the defined API), it is far from clear the occasions on which Level 2 should be used. This paper has explored some of the scenarios in which applications cannot be easily implemented at Level 1 and, therefore, Level 2 support is required. In doing so, we have found no redundant features of Level 2. For each feature (only available at Level~2) we have presented good examples that require use of that feature.

Our studies also reveal some deficiencies in the features provided at Level 2. The lack of convenient support for terminating managed threads is not controversial and is probably just an omission in the current SCJ specification.

It could  be argued that the inability to set a deadline on the transition between missions is not necessary for safety-critical systems as static analysis should have determined whether deadlines can be met. However, we note that SCJ does support detection of deadline misses on managed schedulable objects at Levels 1 and 2.

The need to enrich the mission sequencer concept to support composibility of timing constraints is, perhaps, very controversial as it requires the monitoring of CPU-time usage. Although this is supported by the POSIX standard via sporadic process servers, we are not aware of any implementation of the approach when the threads within the process can execute in parallel.

SCJ support at Level 2 needs to be more complex than at Level 0 or Level 1. The protocol that supports the termination of missions and their sequencers is more complex than is necessary and allows programs to break through the mission hierarchy in an uncontrolled fashion. We have proposed simplifications of the protocol that reinforces the hierarchical nesting of mission sequencers.

\section{Acknowledgements}

This research reported in this paper is funded by the UK EPSRC under grant EP/H017461/1.
Wellings is a member of the Java Community Process JSR 302 Expert Group, which is tasked with developing the Safety-Critical Java Specification. We would like to thank the other members of the Expert Group for their contributions and feedback on some of the ideas expressed in this paper. No new primary data were created during this study.

\bibliographystyle{abbrv}
\bibliography{sigproc}

\end{document}